\documentclass[9pt,letterpaper]{article}

\usepackage[utf8]{inputenc}
\usepackage{amsmath}
\usepackage{amsfonts}
\usepackage{amssymb}
\usepackage{graphicx}
\usepackage{color}

\usepackage[normalem]{ulem}

\usepackage{epstopdf}
\usepackage{hyperref}
\usepackage{cite}

\usepackage{authblk}
\usepackage{sidecap}

\newcommand{\im}{\mathrm{i}}
\newcommand{\E}{{\bf E}}
\newcommand{\EE}{{\bf E}}
\newcommand{\DD}{{\bf D}}
\newcommand{\HH}{{\bf H}}
\newcommand{\PP}{{\bf P}}

\newcommand{\uu}{{\bf u}}

\newcommand{\e}{\varepsilon}
\newcommand{\w}{\omega}

\newcommand{\xx}{{\bf x}}
\newcommand{\rr}{{\bf r}}
\newcommand{\qq}{{\bf q}}
\newcommand{\kk}{{\bf k}}
\newcommand{\me}{{m}}
\newcommand{\wpl}{{\omega_{p}(0)}}
\newcommand{\wplq}{\omega_p}
\newcommand{\wplGr}{\wplq^{Gr}}

\newcommand{\wplTwo}{{\omega_p^{2D}} }

\newcommand{\rhoe}{\rho_{ext}}
\newcommand{\rhoi}{\rho_{ind}}
\newcommand{\rhot}{\rho_{tot}}
\newcommand{\vc}{\upsilon_c}

\providecommand{\enr}[1]{{\epsilon_{#1}}}%
\providecommand{\enrG}[2]{ \epsilon_{#1}^{#2} }%

\providecommand{\IM}[1]{{\text{Im}[{#1}]}}%
\providecommand{\RE}[1]{{\text{Re}[{#1}]}}%
\newcommand{\vPsi}{{\bf\Psi}}
\newcommand{\DP}{k_0\sqrt{\e_x}}

\newcommand{\nsp}{n_{sp}}

\title{Graphene and active metamaterials: theoretical methods and physical properties}

\author[1,2,*]{Marios Mattheakis}
\author[2]{Giorgos P. Tsironis}
\author[1,3]{Efthimios Kaxiras}
\affil[1]{School of Engineering and Applied Sciences, Harvard University, Cambridge, Massachusetts 02138, USA}
\affil[2]{Department of Physics, University of Crete, Heraklion 71003, Greece}
\affil[3]{Department of Physics, Harvard University, Cambridge, Massachusetts 02138, USA}
\affil[*]{Email address: mariosmat@g.harvard.edu}
\date{}

\begin{document}
\maketitle

\section*{Abstract} 
\noindent The interaction of light with matter has triggered the interest of scientists for long time. The area of plasmonics emerges in this context through the interaction of light with valence electrons in metals.  The random phase approximation in the long wavelength limit is used for analytical investigation of plasmons in three-dimensional metals, in a two-dimensional electron gas and finally in the most famous two-dimensional semi-metal, namely graphene.  We show that plasmons in bulk metals as well as in a two-dimensional electron gas   originate from classical laws, whereas, quantum effects appear as non-local corrections. On the  other hand, graphene plasmons are purely quantum modes and, thus, they would not exist in a “classical world”.  Furthermore, under certain circumstances, light is able to couple with plasmons on metallic surfaces, forming a surface plasmon polariton, which is very important in nanoplasmonics  due to its subwavelength nature. In addition, we outline two applications that complete our theoretical investigation. Firstly, we examine how the presence of gain (active) dielectrics affects surface plasmon polariton properties and we find that there is a gain value for which the metallic losses are completely eliminated resulting to lossless plasmon propagation. Secondly, we combine monolayers of graphene in a periodic order and construct a plasmonic metamaterial that provides tunable wave propagation properties, such as epsilon-near-zero behavior, normal and negative refraction.

\noindent
{\bf Keywords:} Random phase approximation, graphene, gain dielectrics, plasmonic metamaterial.

\section{Introduction}
The interaction of light with matter has triggered the interest of scientists for long time. 
The area of plasmonics emerges in this context through the interaction of light with electrons in metals while a plasmon is
 the quantum of the induced electronic collective oscillation. In three-dimensional (3D) metals as well as in a two-dimensional electron gas (2DEG), the plasmon arises classically through a depolarized electromagnetic field generated through Coulomb long range interaction of valence electrons and crystal ions \cite{kittel}. Under certain circumstances, light  is able to couple with plasmons on metallic surfaces, forming a  surface plasmon polariton (SPP) \cite{maier,mariosPT, pitarke}. The SPPs are  very important  in nanoplasmonics and  nanodevices, due to their subwavelength nature, i.e. because their spatial scale is smaller than that of corresponding free electromagnetic modes. In addition to classical plasmons, purely quantum plasmon modes exist in graphene,  the famous two-dimensional (2D) semi-metal.  Since we need the Dirac equation to describe the electronic structure of graphene, the resulting plasmons  are purely quantum objects \cite{grapheneReview,prb34_1986, jablanPRB, graphenePlasmonicsBook}.  As a consequence, graphene is quite special from this point of view, possessing exceptional optical properties,  such as ultra-sub-wavelength plasmons stemming from the specifics of the
 light-matter interaction \cite{mariosENZ,graphenePlasmonicsBook, grapheneReview, basovNature2012}.

In this Chapter, we present basic properties of plasmons,  both from a classical standpoint but also quantum mechanically using the random phase approximation approach.  Plasmons in 3D metals  as well as in 2DEG  originate from  classical laws, whereas, quantum effects appear as non-local corrections \cite{collectiveBook,isihara,kono}. In addition, we point out the fundamental differences between volume (bulk), surface and two-dimensional plasmons. We show that graphene plasmons are a purely quantum phenomenon and that they would not exist in a ``classical world''. We then outline  two applications that complete our theoretical investigation. Firstly, we examine how the presence of gain (active) dielectrics affects SPP properties and we find that there is a gain value for which the metallic losses are completely eliminated resulting to lossless SPPs propagation \cite{mariosPT}. Secondly, we combine monolayers of graphene in a periodic order and construct a plasmonic metamaterial that  provides tunable wave propagation properties, such as  epsilon-near-zero behavior, normal and negative refraction \cite{mariosENZ}.

\section{Volume and surface plasmons in three dimensional metals}
\subsection{Free collective oscillations: Plasmons }
Plasma  is a medium with equal concentration of positive and negative charges, of which at least one charge type is mobile \cite{kittel}. In a classical approach, metals are considered to form a plasma made of ions and electrons. The latter are only the valence electrons that do not interact with each other forming an ideal negatively charged free electron gas  \cite{kittel,psaltakis}. The  positive  ions, i.e. atomic nuclei,  are uniformly distributed  forming a constant  background of positive charge. The background positive charge is considered to be fixed in space, and, as a result, it does not response to any electronic fluctuation or any external field while the electron gas is free to move. In  equilibrium, the electron density (plasma sea) is also distributed uniformly at any point preserving the overall electrical neutrality of the system.  
Metals  support free and collective longitudinal charge oscillation with well defined natural frequency, called the plasma frequency $\wplq$.  The quanta of these charge oscillations are  \textit{plasmons}, i.e.  quasi-particles with energy  $E_p=\hbar \wplq$, where $\hbar$ is the reduced Plank constant. 

\begin{figure}
\centering
\includegraphics[scale=.22]{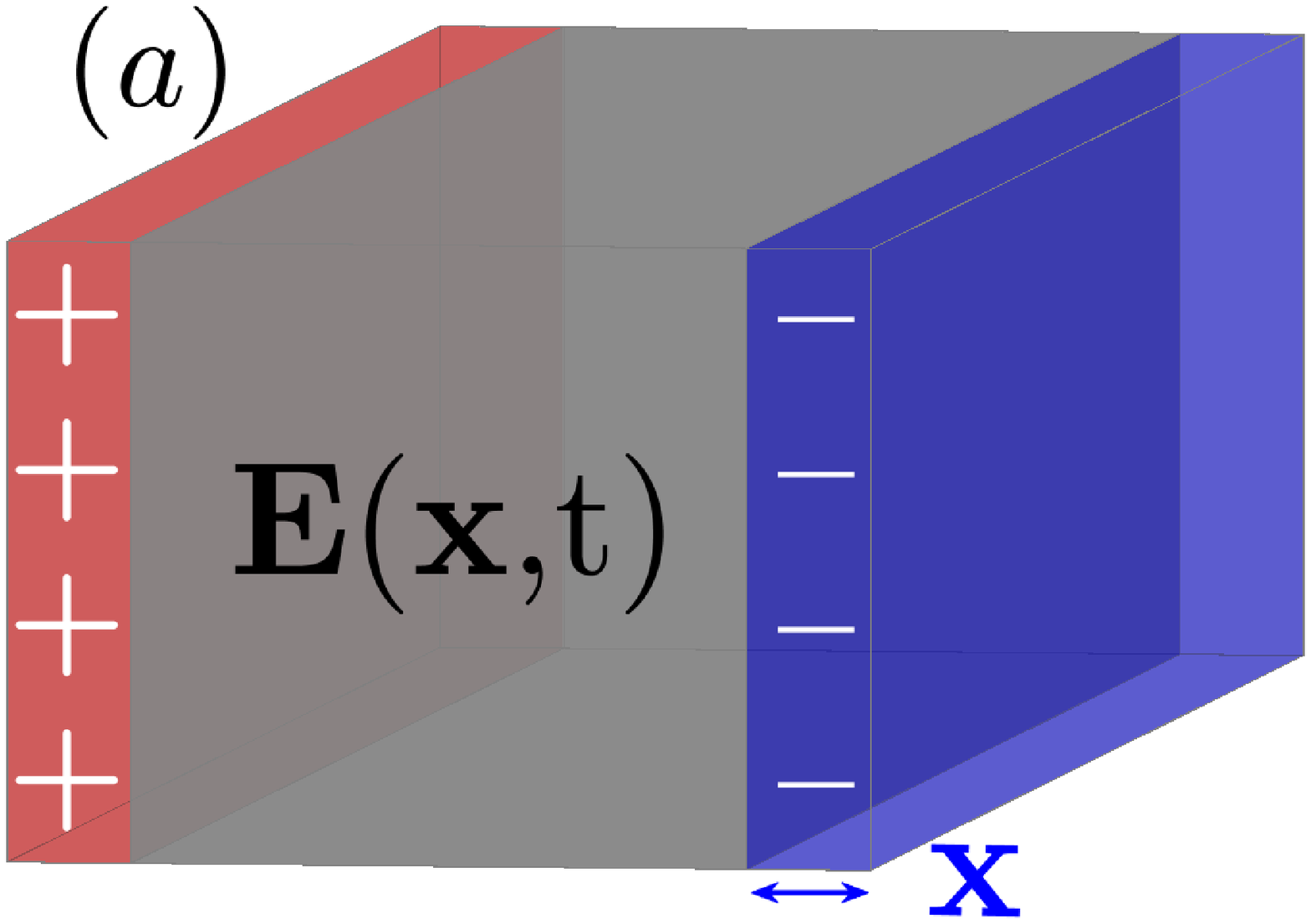}
\includegraphics[scale=.2]{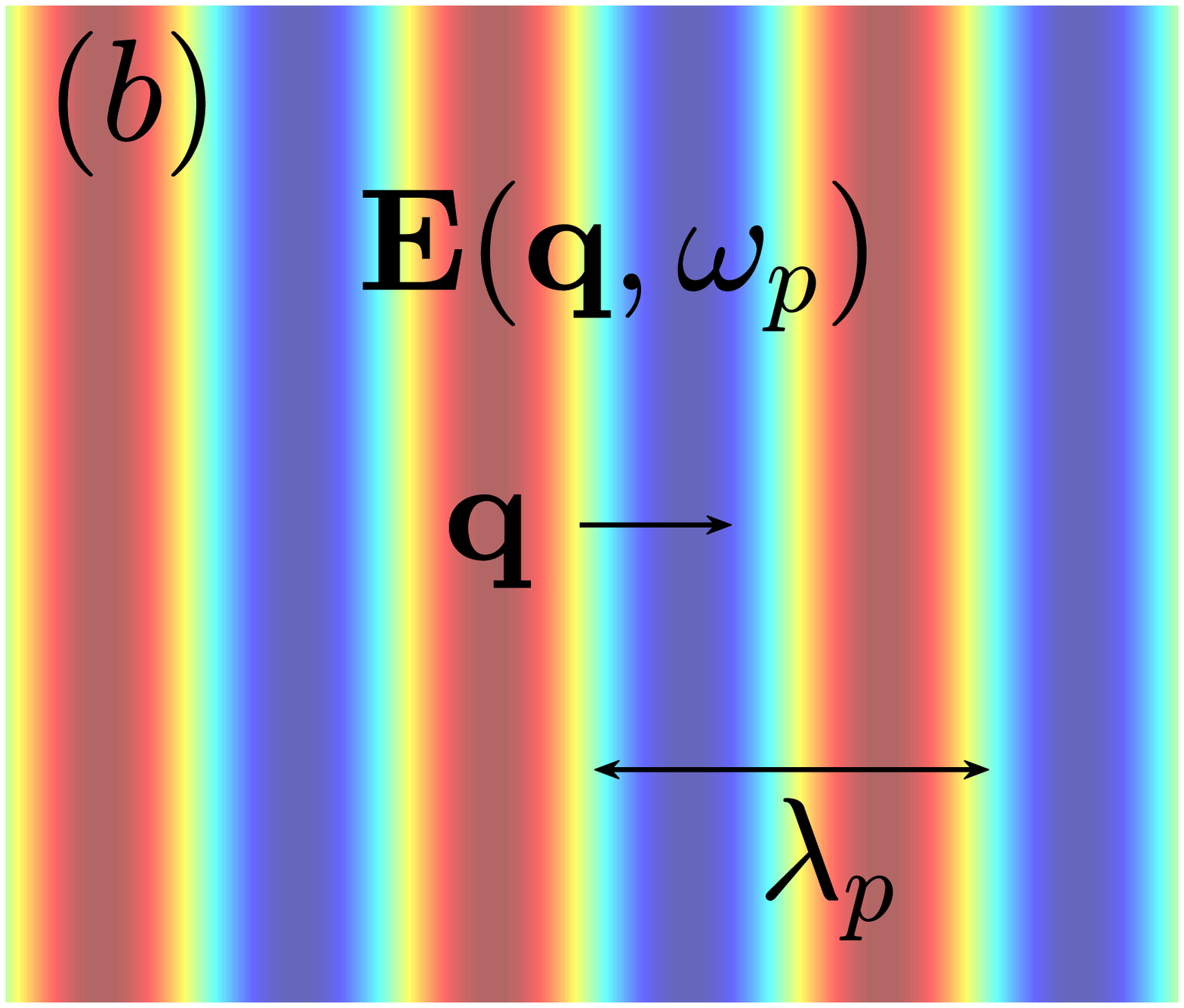}
\caption{(a) A charge displacement is established by displacing uniformly a slab of electrons a small distance $\xx$, creating a polarized electric field in the solid.  (b) A plasma longitudinal oscillation electric field in the bulk of a solid. The arrow indicates the direction of displacement of electrons and of the wavevector $\qq$, while the double faced arrow shows the plasmon wavelength $\lambda_p$.} \label{fig:chargeDisplacement}
\end{figure}

We assume a plasma model with electron (and ion) density $n$. A uniform  charge imbalance $\delta n$ is established in the plasma by displacing uniformly  a zone  of electrons (e.g. a small slab in Cartesian coordinates) by a small distance   $ \xx$ (Fig. \ref{fig:chargeDisplacement}). The uniform displacement implies that all electrons oscillate in phase\cite{maier}; this is compatible with a  long wavelength approximation ($\lambda_{p}/\alpha \rightarrow \infty$, where $\lambda_p$ is the plasmon wavelength and $\alpha$ is the crystal lattice constant); in this  case  the associated wavenumber $|\qq|$ (Fig. \ref{fig:chargeDisplacement}(b))  is very small compared with Fermi wavenumber $k_F$, viz. $q/k_F \rightarrow 0$ \cite{grapheneReview}. Longitudinal oscillations including finite wave vector $\qq$  will be taken into account  later  in the context of quantum mechanics. The immobilized ions form a constant charge density indicated by $e n$, where $e$ is  the   elementary charge.  Let   $\xx(t)$ denote the position of the displaced electronic slab  at time $t$ with charge density given by $-e\delta n(t)$. Due to the electron displacement, an excess positive charge density,  is created that is equal to  $e\delta n(t)$, which  in equilibrium, $\delta  n=0$, reduces to zero. Accordingly, an electric field is generated and interacts with the positive background via Coulomb interaction, forcing the electron cloud to  move as a whole with respect to the immobilized ions, forming an electron density oscillation, i.e. the plasma oscillation. The polarized electric field is determined by the first Maxwell equation as 
\begin{equation}
\label{eq:maxwell1}
\nabla \cdot \E =4 \pi e \delta n,
\end{equation}
in CGS units\footnote{For  SI units we make the substitution $1/ \e_0=4\pi$.}.  The  displacement $\xx(t)$ in the electronic gas produces an electric current density ${\bf J}=-e (n+\delta n)  {\bf \dot{x}}\approx -e n  {\bf \dot{x}}$ (since $\delta n/n \rightarrow 0$), related to the electron charge density  via the continuity equation $\nabla \cdot {\bf J} = -e\partial_t \delta n$. 
%
After integration in time we obtain
\begin{equation}
\label{eq:xx}
 \delta n=n \nabla \cdot \xx 
\end{equation}
Combining the equations (\ref{eq:maxwell1}) and (\ref{eq:xx}), we find the electric field that is induced by the electron charge displacement, i.e.
\begin{equation}
\label{eq:elecField}
\E = 4\pi e n \xx.
\end{equation}
Newtonian mechanics states that an electron with mass $\me$ in an electric field $\E$ obeys the equation $\me \ddot{\xx} = -e \E$,  yielding finally  the equation of motion 
\begin{equation}
\label{eq:motion}
\me \ddot{\xx} +4\pi e^2 n \xx=0,
\end{equation}
indicating that electrons form a collective oscillation  with plasma frequency
\begin{equation}
\label{eq:wp}
\wpl=\sqrt{\frac{4\pi e^2 n }{\me}}.
\end{equation}
where $ \wpl\equiv \wplq( \qq=0 )$.  The energy {$E_p=\hbar\wplq$}  is the minimum energy necessary for exciting a plasmon. Typical values of plasmon energy $E_p$ at metallic densities are in the range $2-20$ eV.

Having shown that an electron gas supports free and collective oscillation modes, we proceed to investigate the dynamical dielectric function $\e(\qq,\omega)$ of the free electron gas. The dielectric function is the response of  the electronic gas to an external electric field and determines  the electronic properties of the solid \cite{kittel, kaxiras, collectiveBook}. 
We consider an electrically neutral homogeneous electronic gas and introduce a weak space-time varying external charge density $\rhoe(\xx,t)$ \cite{psaltakis}. Our goal is to investigate the longitudinal response of the system as a result of  the external perturbation. In  free space, the external charge density produces an  electric displacement field $\DD(\xx,t)$ determined by the divergence relation  $\nabla \cdot \DD=4\pi \rhoe$. Moreover, the system responds and generates additional charges (induced charges) with density $\rhoi(\xx,t)$ creating a polarization field  $\PP(\xx,t)$ defined  by the expression $\nabla \cdot \PP=-\rhoi$ \cite{kittel}. Because of the polarization, the total charge density inside the electron gas will be $\rhot=\rhoe+\rhoi$, leading to the screened electric field $\E$,  determined by  $\nabla \cdot \E=4\pi\rhot$.   The fundamental relation $\DD=\E+4\pi \PP$ is derived after combining  the aforementioned field equations.

The dielectric function is introduced as the linear optical response of the system. According  to the linear response theory and taking into account the non-locality in time and space \cite{maier,psaltakis}, the total field depends linearly on the external field, if the latter is weak. In the most general case we have 
\begin{equation}
\label{eq:LRT}
\DD(\xx,t)=\int d\xx'\int_{-\infty}^\infty dt' \e(\xx-\xx', t-t') \E(\xx',t'),
\end{equation}
where we have implicitly assumed that all length scales are significantly larger than the crystal lattice, ensuring homogeneity. Thence, the response function depends only on the differences between spatial and temporal coordinates \cite{maier,graphenePlasmonicsBook}.
In Fourier space the convolutions turn into multiplications and  the fields are decomposed  into individual plane-wave components of the wavevector $\qq$ and angular frequency $\w$.  Thus,  in the Fourier domain the equation (\ref{eq:LRT}) reads
\begin{equation}
\label{eq:DeE}
\DD(\qq,\w)=\e(\qq,\w)\E(\qq,\w).
\end{equation}
For notational convenience, we designate the Fourier transformed quantities with the same symbol as the original while they differ in the dependent variables. The Fourier transform of an arbitrary field  ${\bf F}(\rr,t)$ is given by ${\bf F}(\rr,t)=\int {\bf F}(\qq,\w)e^{\im(\qq\cdot \rr-\w t)}d\qq dt $  where $\omega$, $\qq$ represent the Fourier transform quantities. 
Hence The Fourier transform of the divergence equations of $\DD$ and $\EE$ yields 
\begin{eqnarray}
\label{eq:FTnablaD}
-\im \qq\cdot \DD(\qq,\w)=4\pi \rhoe(\qq,\w) \\
\label{eq:FTnablaE}
-\im \qq\cdot \EE(\qq,\w)=4\pi \rhot(\qq,\w).
\end{eqnarray}
In longitudinal oscillations the electron displacement field  is in the direction of $\qq$ (Fig. \ref{fig:chargeDisplacement}(b)), thus,   $\qq\cdot\DD=qD$ and $\qq\cdot\EE=qE$, where  $D(\qq,\w)$ and $E(\qq,\w)$ refer to longitudinal fields. Combining the equations (\ref{eq:DeE}), (\ref{eq:FTnablaD}), (\ref{eq:FTnablaE}) yields
\begin{equation}
\label{eq:charge_e}
\rhot(\qq,\w)=\frac{\rhoe(\qq,\w)}{\e(\qq,\w)}.
\end{equation}
Interestingly enough, in the absence of external charges, $\rhoe(\qq,\w)=0$, the equation (\ref{eq:charge_e}) states that non-zero amplitudes of charge oscillation exist, i.e. $\rhot(\qq,\w) \neq 0$, under the condition
\begin{equation}
\label{eq:plasmonCondition}
\e(\qq,\w)=0.
\end{equation}
In other words, in the absence of any external perturbation, free collective charge oscillations exist with dispersion relation $\omega(\qq)$ that satisfies the condition (\ref{eq:plasmonCondition}). These are  plasmon modes and consequently the equation (\ref{eq:plasmonCondition}) is referred as \textit{plasmon condition}. Furthermore, the condition  (\ref{eq:plasmonCondition}) leads to $\EE=-4\pi \PP$, revealing that at  plasmon frequencies the electric field is a pure depolarization field \cite{maier, kittel}. 

We note that  due to their longitudinal nature,  plasmon waves  cannot couple to any transverse wave such as electromagnetic waves; as a result volume plasmons cannot be excited by light. On the other hand,  moving charged particles can be used for exciting plasmons.  For instance, an electron beam passing through a thin metal excites plasmons by transferring part of its energy  to the plasmon excitation. As a result,  plasmons do not decay directly via electromagnetic radiation but  only through energy transfer to electron-hole excitation (Landau damping)  \cite{maier, psaltakis, graphenePlasmonicsBook}.


\subsection{Dynamical Dielectric Function}
Based on the plasmon condition (\ref{eq:plasmonCondition}), the problem has been reduced in the calculation of the dynamical dielectric function $\e(\qq,\w)$. Further investigation of $\e(\qq,\w)$  reveals the plasmon dispersion relation as well as the Landau damping regime, i.e. where plasmons decay very fast exciting electron-hole pairs \cite{graphenePlasmonicsBook}. 
Classically, in the long wavelength limit, the dielectric response $\e(0,\w)$ can be calculated in the context of the plasma model \cite{kittel,collectiveBook}.  Let us consider the  plasma model of equation (\ref{eq:motion}) subjected to a weak and harmonic time-varying external field $\DD(t)=\DD(\w) e^{-\im \w t}$; the equation (\ref{eq:motion}) is  modified to read 
\begin{equation}
\label{eq:motionD}
\me \ddot{\xx}(t) +4\pi e^2 n \xx(t)=-e\DD(t).
\end{equation}
Assuming also  a harmonic in time electron displacement, i.e. $\xx(t)=\xx(\w)e^{-\im \w t}$,  the Fourier transform of  equation (\ref{eq:motionD}) yields 
\begin{equation}
\label{eq:motionDFourier}
\left(-\me \omega^2  +4\pi e^2 n \right) \xx(\qq,\omega)=-e\DD(\qq,\omega).
\end{equation}
Introducing  the equation (\ref{eq:elecField}) in (\ref{eq:motionDFourier}) and using  the relation (\ref{eq:DeE}), we derive  the spatially local  dielectric response
\begin{equation}
\label{eq:drudeEps}
\e(0,\omega)= 1-\frac{\wpl^2}{\omega^2} ,
\end{equation}
where the plasma frequency $\wpl$ is defined in equation (\ref{eq:wp}). Equation (\ref{eq:drudeEps}) verifies that the plasmon condition (\ref{eq:plasmonCondition}) is satisfied at the plasma frequency. The dielectric function (\ref{eq:drudeEps}) coincides with  the Drude model permittivity.

Further investigation of the dynamical dielectric function can be performed  using  quantum mechanics.  An explicit form of $\e(\qq,\w)$ including screening effect  has been evaluated in the context of the \textit{random phase approximation} (RPA) \cite{psaltakis, isihara, kono, graphenePlasmonicsBook} and given by
\begin{equation}
\label{eq:RPAeps}
\e(\qq,\w)=1 - \vc(\qq) \chi_0(\qq,\w)
\end{equation} 
where $v_c(\qq)$ is the Fourier transform of the Coulomb potential and $\chi_0(\qq,\w)$ is the polarizability function, known as Lindhard formula \cite{psaltakis, isihara, kono, graphenePlasmonicsBook}. The Coulomb potential in two and three dimensions, respectively, reads 
\begin{equation}
\label{eq:coulomb}
\vc(\qq)=\begin{cases} ~\frac{2\pi e^2}{|\qq| {\e_b}}  ~ ~\quad \text{(2D)}\\[1.5ex]
~ \frac{4\pi e^2}{|\qq|^2 {\e_b}}  ~\quad~ \text{(3D)}
\end{cases}
\end{equation}
{where $\e_b$ represents the background lattice dielectric constant of the system.} 

{In RPA approach, the dynamical conductivity $\sigma(\qq,\w)$  reads  
\cite{graphenePlasmonicsBook}
\begin{equation}
\sigma=\frac{\im \w e^2}{q^2} \chi_0(\qq,\w),
\end{equation}
revealing the fundamental relation between $\e(\qq,\w)$ and $\sigma(\qq,\w)$ that also depends on system dimensions; we have finally
\begin{equation}
\label{eq:sigmaEps}
\e(\qq,\w)=1+   \im\frac{ q^2 v_c }{\w  e^2 } \sigma(\qq,\w).
\end{equation}
}

In the random phase approximation the most important effect of interactions is that they produce electronic screening, while the electron-electron interaction is neglected. The polarizability of a  non-interacting electron gas is represented by Lindhard formula as follows
\begin{equation}
\label{eq:lindhard}
\chi_0(\qq,\w)=-\frac{2}{V}\sum_\kk \frac{ f(\enr{\kk+\qq})-f(\enr{\kk})}{\hbar\w-\left(\enr{\kk+\qq} -\enr{\kk}\right) + \im\hbar \eta}
\end{equation}
where the factor 2  is derived by spin degeneracy (summation over the two possible values of spin $\mathrm{s}=\uparrow,\downarrow$) \cite{psaltakis, kono, graphenePlasmonicsBook}. The summation is over all the wavevectors $\kk$, $V$ is the volume, $\im \hbar\eta$ represents a small imaginary number to be brought to zero after the summation and $\enr{\kk}$ is the kinetic energy for the wave vector $\kk$.  The carrier distribution $f$ is given by Fermi-Dirac distribution $f(\enr{\kk})=\left( \exp[\beta(\enr{\kk} -\mu)] +1 \right)^{-1}$, where $\mu$ is the chemical potential and $\beta=1/k_B T$ with  the Boltzmann's constant denoted by $k_B$ and $T$ is the absolute temperature. Equation (\ref{eq:lindhard}) describes processes in which a particle in state $\kk$, which is occupied with probability $f(\enr{\kk})$, is scattered into state $\kk+\qq$, which is empty with probability $1-f(\enr{\kk+\qq})$.  The equations (\ref{eq:RPAeps})-(\ref{eq:lindhard}) consist the basic equations for a detailed investigation of  charge density fluctuations and the screening effect, electron-hole pair excitation and plasmons.  With respect to condition (\ref{eq:plasmonCondition}), the roots of equation (\ref{eq:RPAeps}) determine the plasmon modes. Moreover, the poles of $\chi_0$ accounts for  electron-hole pair excitation  defining the plasmon damping regime \cite{psaltakis, isihara, kono}.

For an analytical investigation we split the summation of equation (\ref{eq:lindhard}) in two parts. We  make an elementary change of variables $\kk+\qq \rightarrow -\kk$, in the term that includes $f(\enr{\kk+\qq})$, and  assume that the kinetic energy is symmetric with respect to the wavevector, i.e.  $\enr{\kk}=\enr{-\kk}$. Therefore, the  formula (\ref{eq:lindhard}) yields 
\begin{equation}
\label{eq:lindhard1b}
\chi_0(\qq,\w)=\frac{2}{V}\left( \sum_\kk \frac{ f(\enr{\kk})}{\hbar z-\left(\enr{\kk+\qq} -\enr{\kk}\right)} - \sum_\kk \frac{ f(\enr{\kk})}{\hbar z +\left(\enr{\kk+\qq} -\enr{\kk}\right) } \right)
\end{equation}
where  $z=\w+\im \eta$. At zero temperature the chemical potential is equal to Fermi energy, i.e. $\mu=E_F$ \cite{psaltakis,collectiveBook, graphenePlasmonicsBook}, and the Fermi-Dirac distribution is reduced to Heaviside step function, thus, $f(\enr{\kk})|_{T=0}=\Theta(E_F-\enr{\kk})$. The kinetic energy of each electron of mass $\me$ in state $\kk$ is given by 
\begin{equation}
\label{eq:kineticParabolic}
\enr{\kk}=\frac{\hbar^2|\kk|^2}{2\me},
\end{equation}
hence
\begin{equation}
\label{eq:difEner}
\enr{\kk+\qq}-\enr{\kk}=\frac{\hbar^2}{2\me}\left( |\qq|^2+2 \kk\cdot \qq \right).
\end{equation} 
At zero temperature, because of the Heaviside step function,   the only terms that survive in summation (\ref{eq:lindhard1b}) are those with $|\kk|<k_F$, where $k_F$ is the Fermi wave-number and related to Fermi energy by equation (\ref{eq:kineticParabolic}) as $k_F=(2\me E_F/\hbar^2)^{1/2}$. Subsequently, we obtain  for the Lindhard formula 
\begin{equation}
\label{eq:lindhard1c}
\chi_0(\qq,\w)=\frac{4}{V} \sum_{|\kk|<k_F} \frac{ \enr{\kk+\qq} -\enr{\kk} }{(\hbar z)^2-(\enr{\kk+\qq} -\enr{\kk})^2}
\end{equation}
Summation turns into integration by using $V^{-1}\sum_{|\kk|}(...) \rightarrow (2\pi)^{-3} \int d^3\kk(...)$, hence
\begin{equation}
\label{eq:lindhardIntegral}
\chi_0(\qq,\w)=\frac{4}{(2 \pi)^3}  \int  d^3 \kk \frac{ \enr{\kk+\qq} -\enr{\kk}}{{(\hbar z)^2-(\enr{\kk+\qq} -\enr{\kk})^2}} 
\end{equation}
where the imaginary part in $z$ guarantees the convergence of the integrals around the poles $\hbar\omega=\pm (\enr{\kk+\qq} -\enr{\kk})$. The poles of $\chi_0$ determine the Landau damping regime where plasmons decay  into electron-hole pairs excitation. In particular, the damping regime is a continuum bounded by the  {limit} values of  $(\enr{\kk+\qq} -\enr{\kk})$;  $\kk$ takes its maximum absolute value $|\kk|=k_F$  and the inner product takes the extreme values $k_F\ {\hat\kk}\cdot\qq=\pm k_F|\qq|$.
\begin{equation}
\label{eq:damping}
\frac{\hbar q}{2 \me}\left(q - 2 k_F  \right) < \w <  \frac{\hbar q}{2 \me}\left(q + 2 k_F  \right),
\end{equation} 
where $q=|\qq|$. The Landau damping continuum (electron-hole excitation regime) is demonstrated in Fig. \ref{fig:dispersionRelation} by the shaded area.

Introducing the relation (\ref{eq:difEner}) into (\ref{eq:lindhardIntegral}) and changing to spherical coordinates $(r,\theta,\phi)$, where $r=|\kk|$ and $\theta$ is the angle between $\kk$ and $\qq$,  we obtain
\begin{equation}
\label{eq:lindhardIntegral3}
\chi_0(q,\w)=\frac{2  k_F^4 q}{ (2\pi)^3 \me z^2} \int_0^{2\pi}d\phi \int_0^{1} dx ~x^2 \int_0^\pi  d\theta\frac{ \left(\frac{q}{k_F}+2 x \cos\theta \right)\sin\theta}{1- \left( \frac{v_F q}{z}\right)^2 \left(\frac{q}{2 k_F} + x \cos\theta \right)^2}.
\end{equation}
where  $x=r/k_F$ is a dimensionless variable  and $v_F=\hbar k_F/\me$ is the Fermi velocity.  In the non-static ($\w\gg v_F q$) and long wavelength $(q\ll k_F)$ limits,  we can expand the integral in a power series of $q$.  Keeping up to  $q^3$ orders, we evaluate the integral (\ref{eq:lindhardIntegral3}) and set the imaginary part of $z$ zero, i.e. $z=\w$. That leads to a third order approximation polarizability function
\begin{equation}
\label{eq:chi3Da}
\chi_0(q,\w)= \frac{k_F^3 q^2}{3 \pi^2 \me \w^2}\left(1+\frac{3  v_F^2 q^2}{5 \w^2} \right),
\end{equation}
which, in turn, yields  the dielectric function by using the formula (\ref{eq:drudeEps}) and the three-dimensional Coulomb interaction (\ref{eq:coulomb}), hence
\begin{equation}
\label{eq:RPAresult3Da}
\e(q\rightarrow 0,\w)=1-\frac{\wpl^2}{\w^2}\left( 1+\frac{3}{5}\left( \frac{v_F q}{\w}\right)^2 \right),
\end{equation}
where {vacuum is assumed as background ($\e_d=1$) and we use  the relation  $k_F=(3\pi^2 n)^{1/3}$\cite{kittel,kaxiras} where  $n$ is the electron density. The result (\ref{eq:RPAresult3Da}) is reduced to simple Drude dielectric function (\ref{eq:drudeEps}) for $q=0$.  

The plasmon  condition (\ref{eq:plasmonCondition}) determines the q-dependent plasmon dispersion relation $\wplq(q)$. Demanding $\e(q,\w)=0$, equation (\ref{eq:RPAresult3Da})  yields  approximately
\begin{equation}
\label{eq:dispersion3D}
\wplq(q) \approx \wpl\left( 1+\frac{3}{10}\left( \frac{v_F q}{\wpl}\right)^2 \right).
\end{equation}
Interestingly enough,   the leading term of plasma frequency (\ref{eq:dispersion3D}) does not include any quantum quantity, such as $v_F$ which appears  as  non-local correction in sub-leading terms. That reveals that plasmons in 3D metals are purely classical modes.  Moreover, a gap, i.e. $\wpl$, appears in the plasmon spectrum of three-dimensional metals.  The plasmon dispersion relation (\ref{eq:dispersion3D}) is shown in Fig. \ref{fig:dispersionRelation}.

In the random phase approximation the electrons do not scatter, i.e. collision between electrons and  crystal impurities are not taken into account. As a consequence, the dielectric function is calculated to be purely real; this is nevertheless an unphysical result as can be seen clearly  at  zero frequency where the dielectric function is not well-defined, i.e. $\e(q,0)=\infty$. The problem is cured by introducing a relaxation time $\tau$ in the denominator of the dielectric function as  follows
\begin{equation}
\label{eq:RPAresult3Db}
\e(q\rightarrow 0,\w)=1-\frac{\wplq^2{(q)}}{\w(\w+\im/\tau)}
\end{equation}
We can prove phenomenologically   expression (\ref{eq:RPAresult3Db}) by using the simple plasma model. In particular, we  modify the equation of motion (\ref{eq:motionD}) to a damped-driven harmonic oscillator by assuming that the motion of electron is damped via collisions occurring with a characteristic frequency $\gamma=1/\tau$ \cite{maier}; this approach immediately leads to the dielectric response (\ref{eq:RPAresult3Db}). Typically values of relaxation time $\tau$ are of the order  $10^{-14}$ s, at room temperature.
 The relaxation time is determined  experimentally. In the presence of $\tau$, the dielectric function (\ref{eq:RPAeps}) is well defined at $\w=0$, where the real part of permittivity has a peak with width $\tau^{-1}$ known as \textit{Drude peak}. Furthermore it can be shown that  equation (\ref{eq:RPAresult3Db}) satisfies the Kramers-Kronig relations (sum rules) \cite{kittel,kaxiras, psaltakis}.

\begin{SCfigure}
  \centering
\includegraphics[scale=0.25]{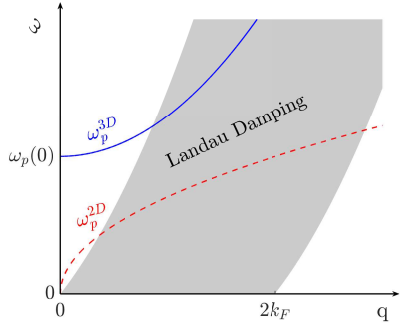}
\caption{Dispersion relation of plasmons in the bulk of three-dimensional solid (blue solid line)  and in two-dimensional electron gas (dashed red curve) plasmons. The shaded region demonstrate the Landau damping regime where plasmons decay to electron-hole pairs excitation.} \label{fig:dispersionRelation}
\end{SCfigure}

\subsection{Surface Plasmon Polaritons}
A new guided collective oscillation mode called  \textit{surface plasmon} arises in the presence of a boundary. Surface plasmon is a surface electromagnetic wave that propagates along an interface  between a conductor (metal) and an insulator (dielectric). This guided mode couples to electromagnetic waves resulting to a polariton. Surface plasmon polaritons (SPPs) occur at frequencies close to but smaller than plasma frequency. These surface modes show exceptional properties for applications of nanophotonics, specifically, they constitute a class of nanophotonics themselves, namely nanoplasmonics. The basic property is the subwavelength nature, i.e. the wavelength of SPPs is smaller than electromagnetic radiation at the same frequency and in the same medium \cite{maier, mariosPT, mariosENZ}.

Let us consider  a waveguide formed by  a planar interface at $z=0$  consisting of two semi-infinite nonmagnetic media (permeability $\mu=1$) with dielectric functions $\e_1$ and $\e_2$ as Fig. \ref{fig:geometry}a denotes.   {The dielectric functions are assumed to be local in space (non $q-$dependent) and non-local in time ($\w$-dependence), hence $\e_{1,2}=\e_{1,2}(\w)$.} Assuming  harmonic in time-dependence in the form ${\bf{u}}(\rr,t)={\bf{u}}(\rr) e^{-\im \w t}$, the Maxwell equations (in CGS units) in the absence of external charges and currents, read
\begin{align}
\label{eq:Maxwell_Ew}
\nabla\cdot (\e_j \EE_j)&=0 &
\nabla\times\EE_j &= {\im k_0} \HH_j\\
\label{eq:Maxwell_Hw}
\nabla\cdot ( \HH_j)&=0 &
\nabla\times\HH_j&=-{\im\e_j k_0} \EE_j
\end{align}
where $k_0=\w/c$ is  the free space  wavenumber and the index $j$  denotes the media as: $j=1$ for $z<0$ and $j=2$  for $z>0$. Combining the equations (\ref{eq:Maxwell_Ew})(\ref{eq:Maxwell_Hw})   the fields are decoupled into two separated Helmholtz equations \cite{maier, pitarke} as
\begin{equation}
\label{eq:helmholtz} 
\left[ \nabla^2+k_0^2\e_j \right]  \left( \begin{array}{c} \EE_j(\rr) \\ \HH_j(\rr)  \end{array} \right) =0
\end{equation}
where $\rr=(x,y,z)$.  For simplicity, let us assume surface electromagnetic waves propagating along one direction, chosen to be the $x$ direction (Fig. \ref{fig:geometry}b), and show no spatial variations in the perpendicular in-plane direction, hence $\partial_y \uu=0$. Under this assumption, we are seeking  electromagnetic waves of the form $\vPsi_j(\rr)=\vPsi_j(z) e^{\im q_j x}$, where $\vPsi_j=(\EE_j,\HH_j)^T$
and $q$  will be the plasmon propagation constant. Substituting the aforementioned ansatz into Helmholtz equation (\ref{eq:helmholtz}), we obtain the guided electromagnetic modes equation \cite{maier}
\begin{equation}
\label{eq:guided}
\left[ \frac{\partial^2}{\partial z^2} +\left(k_0^2 \e_j-q_j^2\right) \right] \left( \begin{array}{c} \EE_j(z) \\ \HH_j(z)  \end{array} \right) =0.
\end{equation}
Surface waves are waves that have been trapped at the interface ($z=0$) and decay exponentially away from it $\left( \vPsi_j(z) \sim e^{-\kappa_j |z|}\text{ for } k_j>0 \right)$ . Consequently, propagating wave solutions along $z$ are not desired. In turn, we  derive to the  \textit{surface wave condition} 
\begin{equation}
\label{eq:surfaceCondition} 
\kappa_j=\sqrt{q^2_j-k_0^2\e_j} \in\mathbb{R}.
\end{equation}
In order to determine the spatial field profiles and the SPP dispersion relation, we need  to find explicit expressions for each field component of $\EE$ and $\HH$. This can be achieved by solving the curl equations (\ref{eq:Maxwell_Ew}) and (\ref{eq:Maxwell_Hw}), which naturally lead to two self-consistent  set of coupled governing equations. Each set corresponds to one of the fundamental polarization, namely, Transverse Magnetic ($p$-polarized waves) and Transverse Electric ($s$-polarized waves), hence
\begin{center} \begin{tabular}{|c|c|}
\hline
{\bf   Transverse Magnetic (TM)} & {\bf Transverse Electric (TE)} \\[0.ex]
\hline 
$\begin{aligned}[t]
\\[-2ex]
 E_{jz}  &= - \frac{q_j}{k_0 \e_j}H_{jy} \\[1ex] 
 E_{jx} &= -\frac{\im}{k_0 \e_j}   \frac{\partial }{\partial z}H_{jy} \\[1ex]
 \frac{\partial^2}{\partial z^2} H_{jy} &- \left(q_j^2-k_0^2 \e_j\right)  H_{jy}  =0  \\[1ex]
\end{aligned}$
&
$\begin{aligned}[t]
\\[-2ex] 
 H_{jz}&= \frac{q}{k_0} E_{jy}  \\[1.0ex] 
 H_{jx}&= \frac{\im}{k_0} \frac{\partial }{\partial z}E_{jy} \\[1ex]
 \frac{\partial^2}{\partial z^2} E_{jy} &- \left(q_j^2-k_0^2 \e_j\right)  E_{jy}  =0  \\[1ex]
\end{aligned}$ 
\\ \hline 
\end{tabular}  \end{center}
We focus on Transverse Magnetic (TM) polarization, in which the magnetic field $\HH$ is parallel to the interface. Since the planar interface extends along  $(x,y)$ plane, the TM fields read $\EE_j=(E_{jx},0,E_{jz})$ and $\HH_j=(0,H_{jy},0)$.  Solving the TM equations  for surface waves, we obtain for each half plane

\vspace{.05\textwidth}
\begin{minipage}{.5\textwidth} %
  \begin{center}
  $z<0 \quad(j=1)$   \end{center}
  \vspace{-.05\textwidth}
\begin{align}
\label{eq:Hy-}
H_y &=~A_1 e^{\im q_1 x} e^{  k_1 z} \\[1ex]
\label{eq:Ex-}
E_x &=-\frac{\im k_1 A_1}{k_0\e_1} e^{\im q_1 x} e^{  k_1 z}  \\[1ex]
\label{eq:Ez-}
E_z &= -\frac{q_1 A_1}{k_0\e_1}e^{\im q_1 x} e^{  k_1 z}  
\end{align}
\end{minipage} %
\begin{minipage}{.5\textwidth} %
    \begin{center}
  $z>0 \quad (j=2)$   \end{center}
    \vspace{-.05\textwidth}
  \begin{align}
\label{eq:Hy+}
H_y &=~A_2 e^{\im q_2 x}e^{ - k_2 z} \\[1ex]
\label{eq:Ex+}
E_x &=~\frac{\im k_2 A_2}{k_0\e_2} e^{\im q_2 x}e^{ - k_2 z}  \\[1ex]
\label{eq:Ez+}
E_z &= -\frac{q_2 A_2}{k_0\e_2}e^{\im q_2 x} e^{ - k_2 z}  
\end{align}
\end{minipage}\vspace{.05\textwidth}
where $k_j$ is related to $q_j$ by equation (\ref{eq:surfaceCondition}). The boundary conditions imply that the parallel to interface components of  electric ($E_x$) and magnetic ($H_y$) fields  must be continuous. Accordingly,  we demand (\ref{eq:Hy-})$=$(\ref{eq:Hy+}) and  (\ref{eq:Ex-})$=$(\ref{eq:Ex+}) at $z=0$, hence we find the system of equations
\begin{equation}
\left(  \begin{array}{cc} e^{\im q_1 x} & -e^{\im q_2 x} \\ \frac{k_1}{\e_1}e^{\im q_1 x} &  \frac{k_2}{\e_2}e^{\im q_2 x} \end{array} \right) \left( \begin{array}{c} A_1 \\ A_2 \end{array} \right) = 0,
\end{equation}
which has a solution only if the determinant is zero. As an outcome, we obtain the so-called \textit{surface plasmon polariton condition}
\begin{equation}
\label{eq:SPcondition}
\frac{k_1}{\e_1}+\frac{k_2}{\e_2}=0.
\end{equation}
The condition (\ref{eq:SPcondition}) states that the interface must consist of  materials with opposite signed permittivities, since surface wave condition requires  the real part of  both $k_1$ and $k_2$ to be non-negative numbers. For that reason, interface between metals and dielectrics may support surface plasmons, since metals show negative permittivity at frequencies smaller than plasma frequency \cite{maier}. Furthermore, boundary conditions demand the continuity of the normal to the interface electric displacement $(D_{jz}=\e_j E_{jz})$ yielding the continuity of the plasmon propagation constant  $q_1=q_2=q$ \cite{pitarke}.
In turn, by combining equations (\ref{eq:surfaceCondition}) with (\ref{eq:SPcondition}) we obtain  the dispersion relation for the surface plasmon polariton
\begin{equation}
\label{eq:SPdispersion}
q(\w) = \frac{\w}{c}\sqrt{\frac{\e_1 \e_2}{\e_1+\e_2}}
\end{equation}
where $\e_{1,2}$ are, in general, complex functions of $\w$.   For a metal-dielectric interface it is more convenient to use the notation $\e_1=\e_d$ and $\e_2=\e_m$ for dielectric and metal permittivity, respectively. In long wavelengths, the SPP wavenumber is close to the light line in dielectric, viz.  $q\simeq k_0 \sqrt{\e_d}$, and the waves are extended over many wavelengths into the dielectrics \cite{maier,pitarke}; these waves are known as Sommerfeld-Zenneck waves  and share similarities with free surface electromagnetic modes \cite{maier}. On the other hand, at the limit $q\rightarrow\infty$, the equation (\ref{eq:SPdispersion}) asymptotically  leads  to the condition 
\begin{equation}
\label{eq:nonretarded}
\e_d+\e_m = 0
\end{equation}
indicating the \textit{nonretarded} surface plasmon limit \cite{pitarke}. In the vicinity of the nonretarded limit,  equation (\ref{eq:surfaceCondition}) yields to $k_j\simeq q \gg k_0$. Furthermore, in the nonretarded limit the phase velocity $v_{ph}=\w/q$ is tending to zero unveiling  the electrostatic nature characterized the surface plasmon \cite{maier,mariosPT}. As a result, at the same frequency  $v_{ph}$   is much smaller than the speed of light and, thus, the SPP wavelength $(\lambda_{sp})$ is always smaller than the light wavelength $(\lambda_{ph})$, i.e. $\lambda_{sp}<\lambda_{ph}$, revealing the subwavelength nature of surface plasmon polaritons \cite{maier,pitarke}. {In addition, due to the fact that SPP phase velocity is always smaller than phase velocity of propagating electromagnetic waves, SPPs cannot radiate and, hence, they are well defined surface propagating electromagnetic waves.} Demanding $q\rightarrow \infty$ in the dielectric function (\ref{eq:RPAresult3Db}), we find the so-called  \textit{surface plasmon frequency $\w_{sp}$}, which is  the  upper frequency limit that SPPs occur
\begin{equation}
\label{eq:wsp}
\omega_{sp}=\sqrt{\frac{\wplq^2}{1+\e_d}-\gamma^2}\simeq \frac{\wplq}{\sqrt{1+\e_d}},
\end{equation}
indicating that SPPs  always occur at frequencies smaller than bulk plasmons. 

If we follow the same procedure for  transverse electric polarized fields, in which the electric field is parallel to interface and the only non-zero electromagnetic field components are $E_y,~H_x,$ and $H_z$, we will find the condition $k_1+k_2=0$ \cite{maier}. This condition is satisfied only for $k_1=k_2=0$ unveiling that  $s$-polarized surface modes do not exist. Consequently, surface plasmon polaritons are always TM electromagnetic waves.

\begin{figure}[ht!]
\centering
\includegraphics[scale=0.267]{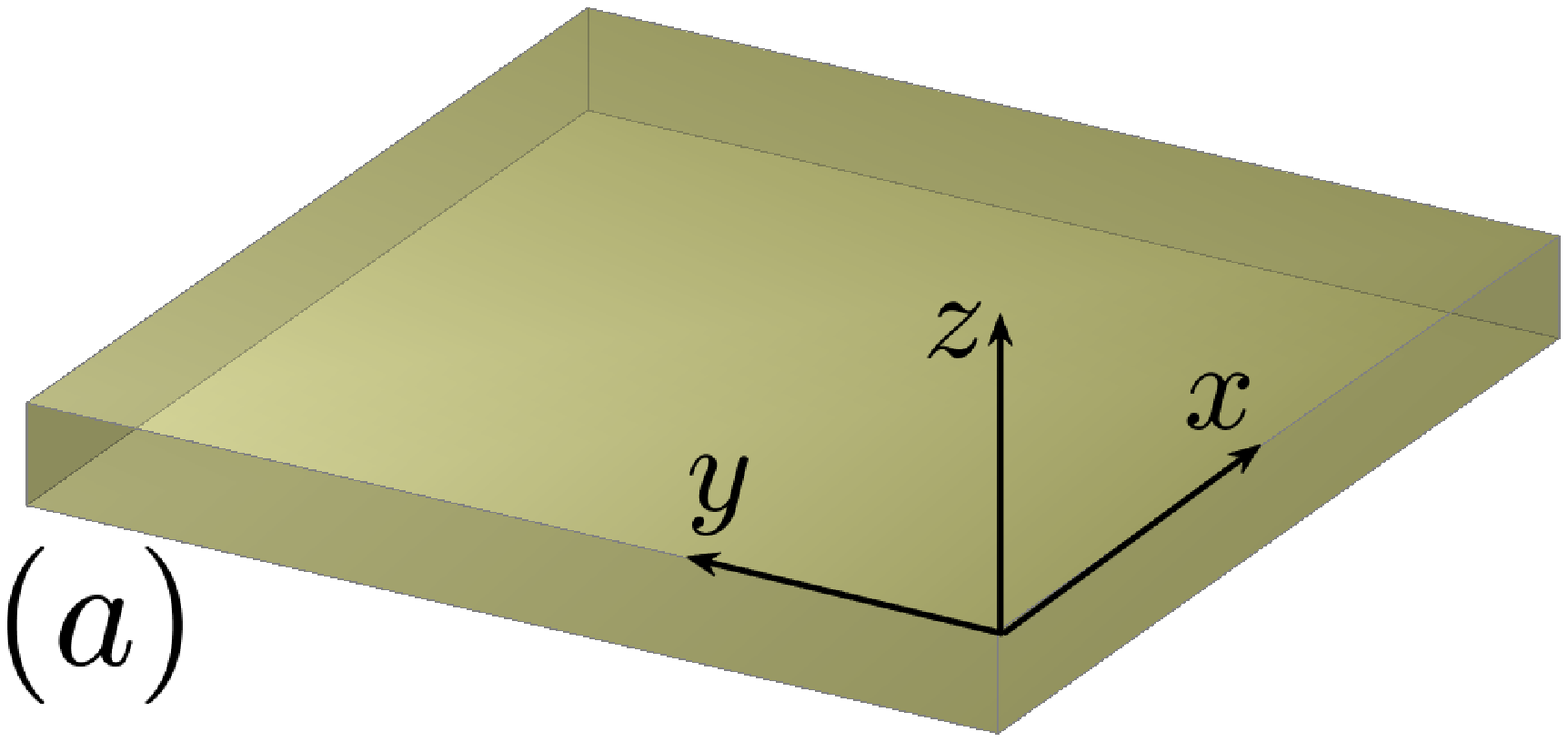}
\includegraphics[scale=0.267]{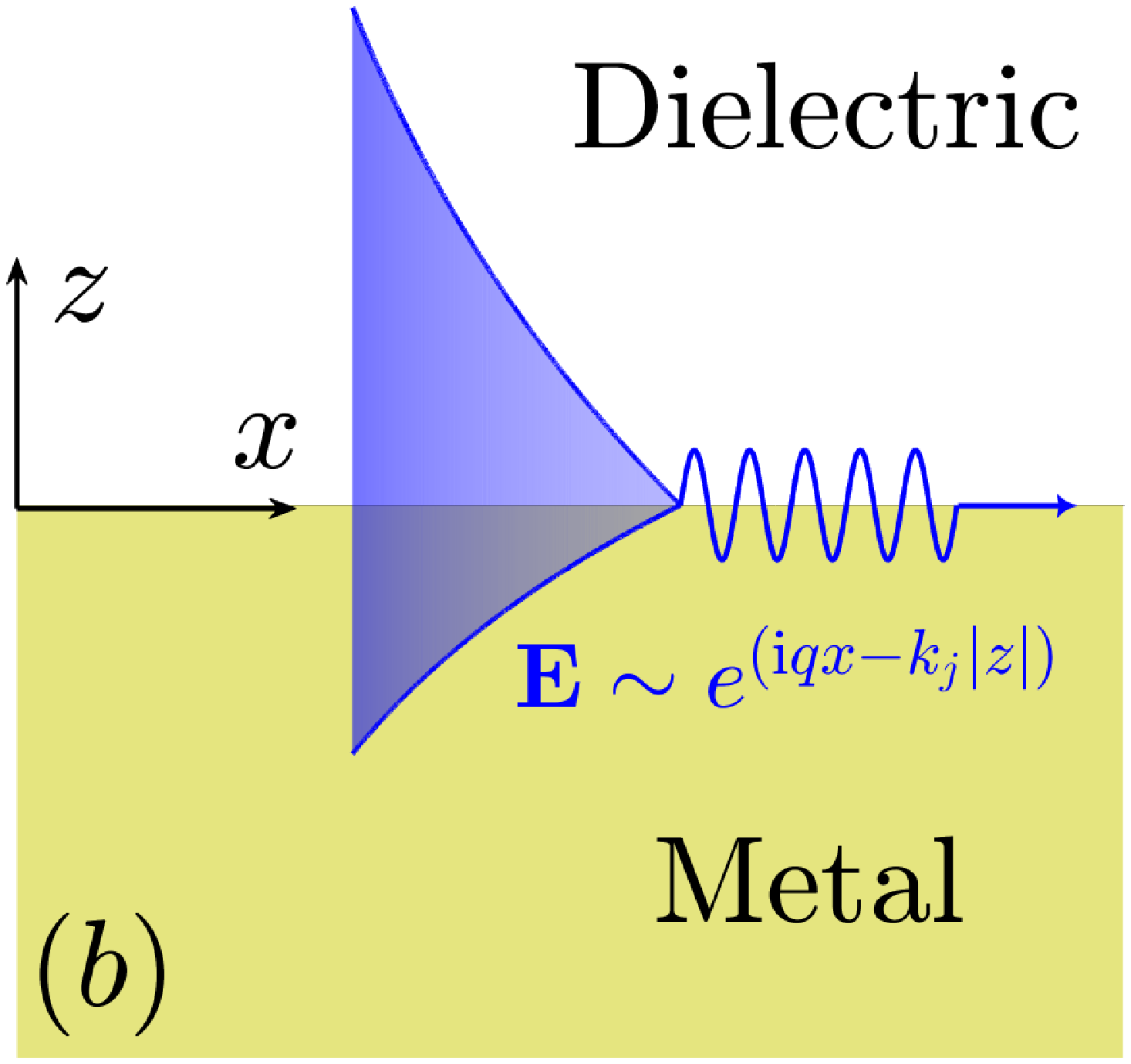}
\caption{A planar interface is formed between a metal and a dielectric where surface plasmon polaritons (SPPs) propagate in (a) three and (b) two dimensional representation. (b) A schematic illustration of the SPP field. }
\label{fig:geometry}
\end{figure}

Due to metallic losses, SPPs decay exponentially along the interface restricting the propagation length. Mathematically speaking,  losses are described by the small imaginary part in the complex dielectric function of metal  $\e_m=-\e_m'-\im\e_m''$, where $\e_m',\e_m'' >0$.  Consequently, the SPPs propagation constant (\ref{eq:SPdispersion}) becomes complex, that is, $q=q'+\im q''$, where the imaginary part accounts for losses of SPPs energy. In turn, the effective propagation length $L$, which shows the rate of change of the energy attenuation of SPPs \cite{maier,mariosPT}, is determined by the imaginary part  $\IM{q}$ as  $L^{-1}={2 \IM{q}}$.

Gain  materials rather than passive regular dielectrics, have been used to reduce the losses in SPP propagation. Gain materials are characterized by a complex permittivity function, i.e. $\e_d=\e_{d}'+\im\e_d''$, with $\e_d',\e_d''> 0$, where $\e_d''$ is a small number compared to $\e_d'$ and accounts for gain. As a result, gain dielectric gives energy to the system counterbalancing the metal losses. We investigate the SPP dispersion relation (\ref{eq:SPdispersion}) in the presence of gain and loss materials, and find an explicit formula for gain $\e_d''$ where  the SPP wavenumber is reduced to real function, resulting to lossless SPPs propagation. In addition, we find an upper limit that values of gain are allowed. In this critical gain the purely real SPPs propagation constast becomes purely imaginary, destroying the SPPs modes. 

The dispersion relation (\ref{eq:SPdispersion}) can also be written as $q=k_0 \nsp$ \cite{mariosPT}, where $\nsp$ is the plasmon effective refractive index given by
\begin{equation}
\label{eq:SPindex}
\nsp=\sqrt{\frac{\e_d \e_m}{\e_d+\e_m}}.
\end{equation}
 We are seeking for a gain $\e_d''$ such that the effective index $\nsp$  becomes real.  Substituting  the complex   function describing the dielectric and  metal into equation (\ref{eq:SPindex}), the function $\nsp$ is written in the ordinary complex form as \cite{mariosPT}
\begin{equation}
\label{eq:nExpanded}
\nsp=  \sqrt{\frac{\sqrt{x^2+y^2}+x}{2}} + \im \ \text{sgn}(y)\sqrt{\frac{\sqrt{x^2+y^2}-x}{2}},
\end{equation}
where $\text{sgn}(y)$ is the discontinuous signum function \cite{mariosPT} and

\begin{minipage}{.5\textwidth} %
\begin{align}
\label{eq:x}
x=\frac{\varepsilon_{d}'|\varepsilon_m|^2-\varepsilon_{m}' |\varepsilon_d |^2}{|\varepsilon_d+\varepsilon_m |^2}
\end{align}
\end{minipage} 
\begin{minipage}{.5\textwidth} %
\begin{align}
\label{eq:y}
 y=\frac{\varepsilon_{d}'' |\varepsilon_m |^2- \varepsilon_{m}'' |\varepsilon_d |^2}{ |\varepsilon_d+\varepsilon_m |^2}
\end{align}
\end{minipage} 
with $|z_*|$ denoting the norm of the complex number $z_*$.  The poles in equations (\ref{eq:x}) and (\ref{eq:y}) correspond to the nonretarted  surface plasmon limit (\ref{eq:nonretarded}).

Considering  the plasmon effective index $\nsp$  in equation (\ref{eq:nExpanded}) in the $(x,\,y)$ plane, we  observe that lossless SPP propagation $(\IM{\nsp}=\IM{q}=0)$, is warranted when the conditions $y=0$ and $x>0$ are simultaneously satisfied. Let us point out that for $y=0$ and  $x<0$, although the imaginary part in equation (\ref{eq:nExpanded}) vanishes due to the signum function, its real part  becomes imaginary, i.e. $\nsp=\im\sqrt{|x|}$, which does not correspond to propagation waves.  Solving the equation (\ref{eq:y}) for $y=0$ with respect to gain $\e_{d}''$ and avoiding the nonretarded limit (\ref{eq:nonretarded}), i.e. $\varepsilon_{d}\neq-\varepsilon_m$,  we obtain two exact solutions \cite{mariosPT} as follows
\begin{equation}
\label{eq:PTgain}
 \varepsilon_{d\pm}'' = \frac{ |\e_m |^2}{2\e_{m}''} \left( 1 \pm \sqrt{1- \left( \frac{2 \e_{d}' \e_{m}''} { |\e_m |^2} \right)^2 } \right).
\end{equation}
Due to the fact that $\e_d$ is real, we read from  equation (\ref{eq:PTgain}) that  \cite{mariosPT}.
\begin{equation}
\label{eq:inequality}
|\e_m|^2 \geqslant 2\e_d'\e_m''.
\end{equation}
Using the inequality (\ref{eq:inequality}), we read for the solution $\e_{d+}$ of (\ref{eq:PTgain}),  that $\e_{d+}'' \geqslant \e_d'$. This is a contradiction since  the $\e_d''$ is defined to be smaller than $\e_d'$. Thus,  $\e_{d+}$  does not  correspond to a physically relevant gain. 

Solving, on the other hand, the equation (\ref{eq:x}) for $x>0$, with respect to the dielectric gain $\e_{d}''$,  we determine a critical value $\e_{c}$ distinguishing the regimes of lossless and prohibited SPP propagation \cite{mariosPT}, namely
\begin{equation}
\label{eq:criticalGain}
\e_c=\e_{d}'\sqrt{\frac{|\e_m|^2}{\e_m'\e_{d}',
}-1},
\end{equation}
hence, equation (\ref{eq:criticalGain}) sets an upper limit in values of gain. The appearance of critical gain can be understood as follows: In equation (\ref{eq:PTgain})  the  gain $\e_{d-}$ becomes equal to critical gain $\e_c$ when $\e_d+\e_m =0$ \cite{mariosPT}, where the last item is the nonretarded limit where $q\rightarrow\infty$. Specifically, the surface plasmon exists when the metal is characterized by the Drude dielectric function of equation (\ref{eq:RPAresult3Db}),  $\e_{d-}''=\e_c$ at $\w=\omega_{sp}$, corresponding  to a maximum frequency\cite{mariosPT}.  

In order to represent the above theoretical findings, we use the dielectric function of equation (\ref{eq:RPAresult3Db}) to calculate the SPP dispersion relation for an interface consisting of  silver with  $\wpl=13.67$ PHz and $\gamma=0.1018$ PHz,  and silica glass with $\e_d'=1.69$ and for gain  $\e_d''=\e_{d-}$  determined by equation (\ref{eq:PTgain}). We represent in Fig. \ref{fig:LosslessDispersion}a the SPP dispersion relation of equation (\ref{eq:SPdispersion}) for lossless case ($\e_d''=\e_{d-}''$), where the lossless gain is denoted by the inset image in Fig.\ref{fig:LosslessDispersion}a. We indicate the real and imaginary of normalized SSP dispersion $q/k_p$ ($k_p\equiv \wplq/c$), with respect to the normalized frequency $\w/\wplq$.  We observe, indeed, that for $\w<\omega_{sp}$ the imaginary part of $q$ vanishes whereas for $\w>\omega_{sp}$ the SPPs wave number is purely imaginary. Subsequently, in the vicinity of $\w=\omega_{sp}$ a phase transition from lossless to prohibited SPPs propagation is expected \cite{mariosPT}.

 We also solve numerically the full system of Maxwell equations  (\ref{eq:Maxwell_Ew}) (\ref{eq:Maxwell_Hw}) in a two-dimensional space for transverse magnetic polarization. The numerical experiments have been performed by virtue of the multi-physics commercial software COMSOL and the frequency $\w$ is confined in the range  $[0.3\wplq,0.75\wplq]$ with the integration step $\Delta \w=0.01\wplq$. In the same range, the lossless gain is calculated by equation (\ref{eq:PTgain}), to be   $[8\cdot10^{-3},8\cdot 10^{-2}]$. For  the excitation of SPPs on  the metallic surface, we use the near-field technique \cite{maier,mariosPT,mariosENZ,basovNature2012}. {For this purpose, a circular electromagnetic source of radius $R=20$nm has been located $100$nm above the metallic surface acting as a point source, since the wavelength $\lambda$ of  EM waves are much larger, i.e. $\lambda>>R$ \cite{maier, mariosPT}.  
{In Fig.\ref{fig:LosslessDispersion}b we  demonstrate, in a log-linear scale, the propagation length $L$, with respect to $\w$, subject in lossless gain $\e_{d-}$ (blue line and open circles).  For the sake of comparison, we plot $L(\omega)$ in the absence of gain (green line and filled circles). The solid lines represent the theoretical predictions obtained by the definition of $L$, whereas the circles indicate numerical results. For the numerical calculations, the characteristic propagation length has been estimated by the inverse of the slope of the Log$(I)$, where $I$ is the magnetic intensity along the interface \cite{maier,pitarke,mariosPT}. The black vertical  dashed line denotes the SPP resonance frequency $\omega_{sp}$, in which the phase transition appears. The graphs in Fig.  \ref{fig:LosslessDispersion}b indicate that in the presence of the lossless gain, SPPs may travel for very long, practically infinite, distances. Approaching the resonance frequency $\omega_{sp}$,  $L$ decreases rapidly leading to a steep phase transition on the SPPs propagation. The  deviations between theoretical and numerical results in Fig.
\ref{fig:LosslessDispersion} for frequencies near and greater than $\omega_{sp}$ are attributed  to the fact that in the regime $\w_{sp}<\w<\wplq$, there are quasi-bound EM modes \cite{maier, mariosPT}, where EM waves are evanescent along the metal-dielectric interface and radiate perpendicular to it. Consequently, the observed EM field for $\w>\w_{sp}$ corresponds to radiating modes \cite{mariosPT}}. 

\begin{figure}[ht!]
\centering
\includegraphics[scale=0.27]{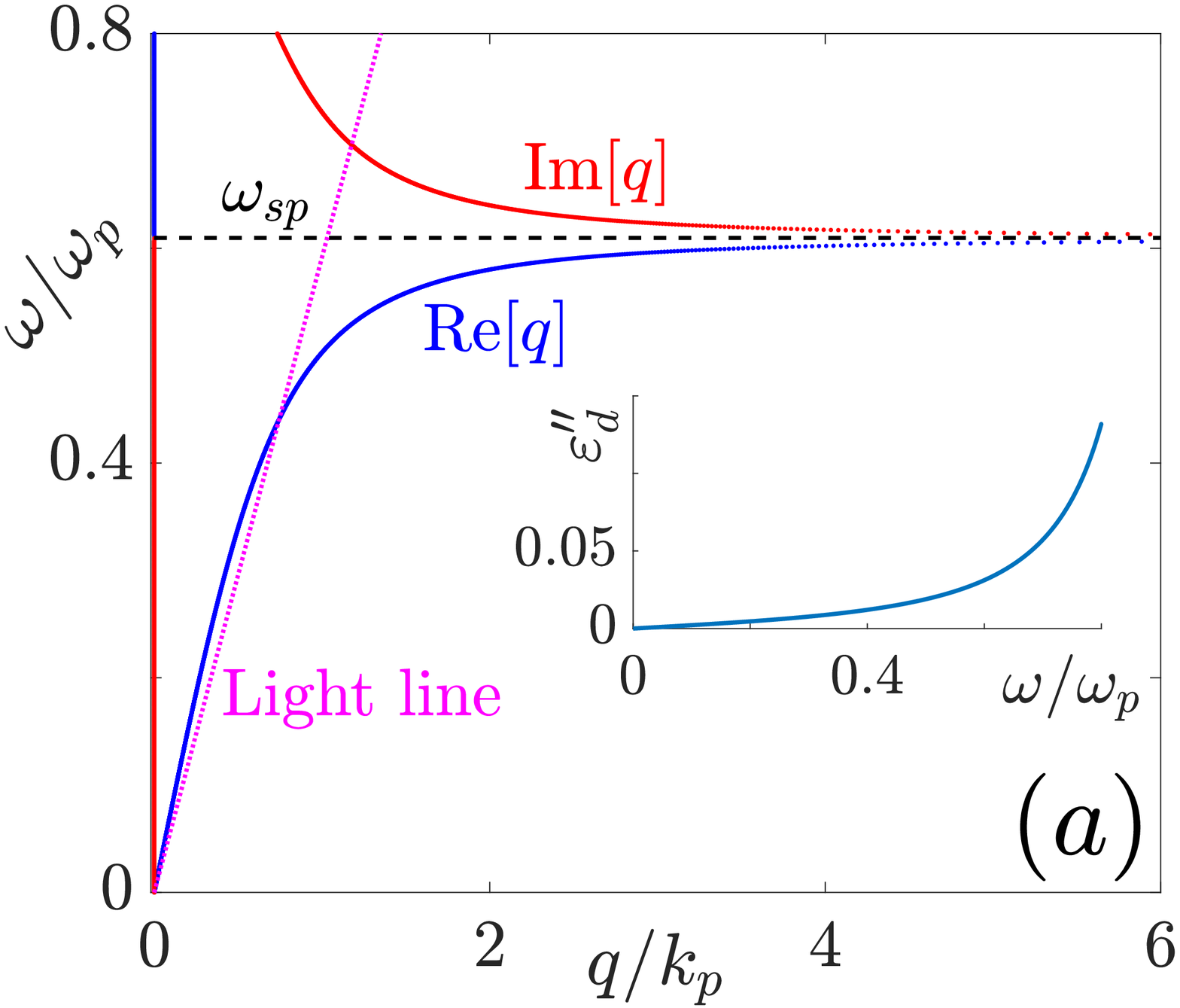}
\includegraphics[scale=0.27]{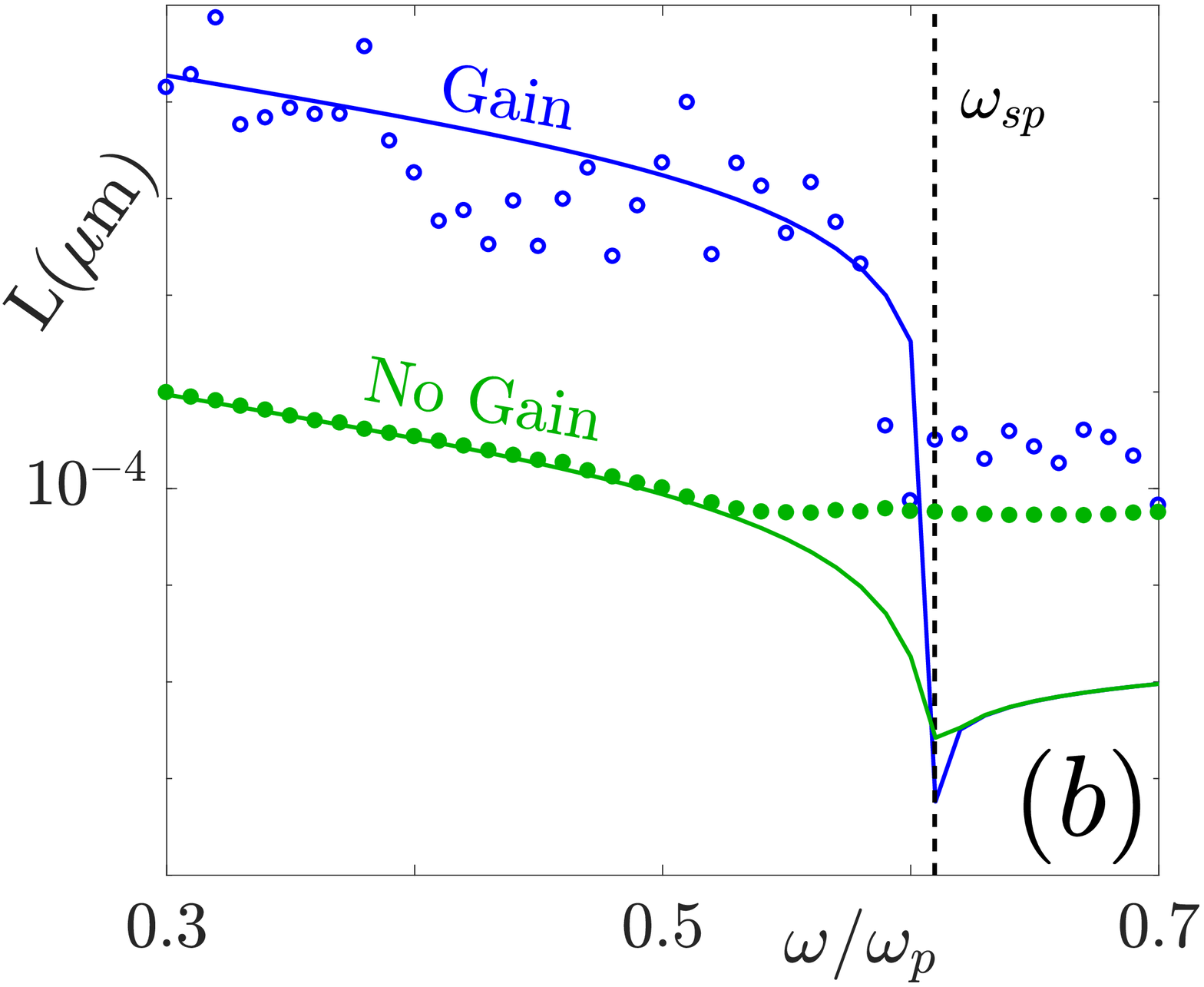}
\caption{(a) The surface plasmon polariton (SPP) dispersion relation $q(\w)$  in the presence of a gain material with gain corresponds to lossless SPP propagation. $\RE{q}$ and $\IM{q}$ are indicated by blue and red line, respectively. The horizontal dashed black line denotes the SPP frequency $(\w_{sp}=0.61\omega_p)$ where an interchanging between  $\RE{q}$ and $\IM{q}$ appears. The dotted magenta line indicates the light line in the dielectric. (Inset) Demonstration of the gain leads to lossless SPP propagation. (b) Theoretical (solid lines) and numerical (circles) prediction of  SPP propagation length $L$ in the presence (blue) and in absence (green) of gain dielectric showing a phase transition that happens at $\w_{sp}$ (vertical dashed black line). Deviations between theoretical and numerical predications for $\w>\w_{sp}$ correspond to quasi-bound EM modes. The $k_p=\wplq/c$ is used as normalized unit of wavenumbers and $\wplq$ as normalized unit for frequencies.}
\label{fig:LosslessDispersion}
\end{figure}

\section{Two-Dimensional Plasmons}
In this section we  investigate plasmons in a two-dimensional electron gas (2DEG), where the electron sea is free to move only in two dimensions, tightly confined in the third. The reduced dimensions of electron confinement and  Coulomb interaction cause crucial differences in plasmons excitation spectrum. For instance, plasmon spectrum in a 2DEG is gapless in contrast with three-dimensional case \cite{kono}. {For the sake of completeness,  we first discuss briefly plasmons in a regular 2DEG characterized by the usual  parabolic dispersion relation  (\ref{eq:kineticParabolic}) for a two-dimensional wavevector $\kk$ lies in the plane of 2DEG. Thence,}  we focus on plasmons in a quite special two-dimensional material, viz. graphene. Graphene is a gapless two-dimensional  semi-metal with linear dispersion relation. The linear energy spectrum is giving the great opportunity to describe graphene with chiral Dirac Hamiltonian for massless spin-$1/2$ fermions \cite{grapheneReview, basovNature2012, graphenePlasmonicsBook}. Furthermore, graphene can be doped with several methods, such as chemical doping \cite{grapheneReview}, by applying an external voltage \cite{basovNature2012} or with lithium intercalation \cite{sharmila}. The doping shifts the Fermi level towards to the conduction bands making graphene a great metal. The advantage to describe graphene electronic properties with massless carriers Dirac equation leads to exceptional optical and electronic properties, like  very high electric conductivity and ultra-sub-wavelength plasmons \cite{jablanPRB, grapheneReview, basovNature2012, graphenePlasmonicsBook}.

\subsection{Dynamical Dielectric Function of 2D metals}

In order to determine the plasmon spectrum of a two-dimensional electron gas, first of all we   calculate the dielectric function in the context of random phase approximation (\ref{eq:RPAeps}) with $v_q$ being the two-dimensional Coulomb interaction of equation (\ref{eq:coulomb}). In the   Lindhard formula  (\ref{eq:lindhard1c}) $V$ and $\kk$ denote a two-dimensional volume and wave-vector, respectively.  Firstly, we investigate a 2DEG described by the parabolic dispersion relation (\ref{eq:kineticParabolic}). The electrons are assumed to occupy a singe band ignoring interband transitions, i.e.  transitions to higher bands. Thus, there is not orbital degeneracy  $(g_v=1)$ resulting to the two-dimensional Fermi wave-number  $k_F=\sqrt{2 \pi n}$, where $n$ is the carrier (electrons) density \cite{kono,prl18_1967}. Turning the summation (\ref{eq:lindhard1c}) into integral by the substitution $V^{-1}\sum_{|\kk|}(...)=(2\pi)^{-2}\int d^2 \kk(...)$, we obtain the Lindhard formula in integral form
\begin{equation}
\label{eq:lindhardIntegral2D}
\chi_0(\qq,\w)=\frac{4}{(2 \pi)^2}  \int  d^2 |\kk| \frac{ \enr{\kk+\qq} -\enr{\kk}}{{(\hbar z)^2-(\enr{\kk+\qq} -\enr{\kk})^2}} 
\end{equation}
The singe particle excitation continuum is still defined by expression (\ref{eq:damping}), since the kinetic energy  is considered to have the same form as in 3D case, even though, the  2D Fermi wavenumber has been modified. Transforming to polar coordinate system $(r,\theta)$ and using the relation (\ref{eq:difEner}), the integral (\ref{eq:lindhardIntegral2D}) reads 
\begin{equation}
\label{eq:lindhardIntegral2Db}
\chi_0(q,\w)=\frac{2  k_F^3 q}{ (2\pi)^2 \me z^2}  \int_0^{1} dx ~x \int_0^{2 \pi}  d\theta\frac{  \frac{q}{k_F}+2 x \cos\theta }{1- \left( \frac{v_F q}{z}\right)^2 \left(\frac{q}{2 k_F} + x \cos\theta \right)^2}
\end{equation}
where $x$ is a dimensionless variable defined as $x=r/k_F$. As previously,  since we we are interested in long wavelength limit ($q \ll k_F$), we   expand the integrand of (\ref{eq:lindhardIntegral2Db})  around $q=0$. Keeping up to first orders of $q$, the integral (\ref{eq:lindhardIntegral2Db}) yields 
\begin{equation}
\label{eq:chi2DEG_1st}
\chi_0(q,\w)=\frac{k_F^2 q^2}{2\pi m \w^2}
\end{equation}
where $z\rightarrow \w$  by sending the imaginary part of $z$ to zero. The dielectric function is determined by the formula of (\ref{eq:RPAeps}) for 2D Coulomb interaction of  (\ref{eq:coulomb}), hence
\begin{equation}
\label{eq:eps2DEG_1st}
\e(q,\w)=1 - \frac{2 \pi n e^2 q}{m \w^2}
\end{equation}
The 2DEG plasmon dispersion relation is determined by (\ref{eq:plasmonCondition}) to be
\begin{equation}
\label{eq:wp2DEG_1st}
\wplTwo(q)=\sqrt{\frac{2 \pi n e^2 q}{m}}
\end{equation}
 related with volume plasmons dispersion relation by $\wplTwo(q)=\wplq\sqrt{q/2}$. In contrast to  three-dimensional electron gas where plasmon spectrum is gapped, in two-dimensional case the plasmon frequency depends on $\sqrt{q}$ making the plasmon spectrum gapless. In Fig.\ref{fig:dispersionRelation}, the 2D plasmon dispersion relation (\ref{eq:wp2DEG_1st}) is demonstrated together with three-dimensional case. Furthermore, it  worths to point out the similarity  between  the plasmon dispersion relation of 2DEG of equation (\ref{eq:wp2DEG_1st}) and SPP of equation (\ref{eq:SPdispersion}), that is, both  show  $\sqrt{q}$ dependence.

Let us now investigate the most special two dimensional electron gas, namely  graphene.  At the limit  {where the excitation energy is small compared to $E_F$,}  the dispersion relation of graphene, viz. the relation between kinetic energy $\enrG{ \kk}{s}$ and momentum ${\bf p}=\hbar \kk$,   is described by two linear bands  as
\begin{equation}
\label{eq:grapheneDispersion}
\enrG{\kk}{s}  ~ = s \hbar v_F |\kk|
\end{equation}
where $s=\pm 1$ indicates the conduction (+1) and valence  (-1) band, respectively, $v_F$ is the two dimensional Fermi velocity which is constant for graphene and equal to $v_F=10^6$m/s \cite{sharmila,prb75_2007,grapheneReview,basovNature2012,graphenePlasmonicsBook}.  Because of valley degeneracy $g_v=2$, the Fermi momentum is modified to read $k_F=\sqrt{2 \pi n/g_v}=\sqrt{\pi n}$   \cite{graphenePlasmonicsBook,prb75_2007}}. The Fermi energy, given by $E_F=\hbar v_F k_F$,  becomes zero in the absence of doping  ($n=0$). As a consequence,  the $E_F$ crosses the point where the linear  valence and conduction bands touch each other, namely at the Dirac point, giving rise to the semi-metal character of the undoped graphene \cite{sharmila,kaxiras,prb75_2007,grapheneReview}. The Lindhard formula of equation (\ref{eq:lindhard}) needs to be generalized to include both intra- and interband transitions {(valley degeneracy)} as well as the overlap of states, hence 
\begin{equation}
\label{eq:LindhardGr}
\chi_0(\qq,\w)=-\frac{g_s g_v}{V}  \sum_{s, s'} \sum_\kk \frac{f(\enrG{\kk+\qq}{s'})-f(\enrG{\kk}{s}) }{\hbar\w-(\enrG{\kk+\qq}{s'} -\enrG{\kk}{s}) + \im\hbar \eta} F_{ss'}(\kk,\kk+\qq)
\end{equation}
where the factors $g_s=g_v=2$ account to spin and  valley degeneracy, respectively. The Lindhard formula has been modified to contain two extra summations ($\sum_{s=-1}^1\sum_{s'=-1}^1$) corresponding to valley degeneracy for the two bands of equation (\ref{eq:grapheneDispersion}). In addition, the overlap of states function $F_{ss'}(\kk,\kk+\qq)$ has been introduced and defined by $ F_{ss'}(\kk,\kk+\qq)=(1+s s' \cos\psi)/2$, where $\psi$ is the angle between $\kk$ and $\kk+\qq$ vectors \cite{prb34_1986,prb75_2007}. The term $\cos\psi$ can be expressed in $|\kk|$, $|\kk+\qq|$ and $\theta$ terms, and subsequently,  the overlap function is written as \cite{graphenePlasmonicsBook}
\begin{equation}
\label{eq:overlapF}
F_{ss'}(\kk,\kk+\qq)=\frac{1}{2}\left( 1 + ss'\frac{|\kk| + |\qq|\cos\theta}{|\kk+\qq|}\right).
\end{equation}
In long wavelength limit,  we approximately obtain
 \begin{equation}
\label{eq:k+q} 
|\kk+\qq|=|\kk|\left(1+ \frac{|\qq|\cos\theta}{|\kk|}+ \frac{|\qq|^2 \sin^2\theta}{2 |\kk|^2}\right). 
 \end{equation}
In this limit, we obtain for the graphene dispersion relation (\ref{eq:grapheneDispersion}) the general form
 \begin{equation}
\label{eq:Grde}
\enrG{\kk+\qq}{s} -\enrG{\kk}{s'} = s\hbar v_F \left( \frac{s-s'}{s}|\kk|+ |\qq|\left(\cos\theta+\frac{|\qq|}{2 |\kk|}\sin^2\theta\right)\right).
\end{equation}
In turn, the plasmon damping regimes   are determined by the poles of polarizability (\ref{eq:LindhardGr}) by substituting the expression (\ref{eq:Grde}).  Due to the valley degeneracy there are two damping regimes corresponding, respectively, to intraband $(s=s')$
\begin{equation}
\label{eq:dampingIntra}
\w < v_F q
\end{equation}
 and interband ($s=-s'$)
 \begin{equation}
\label{eq:dampingInter}
v_F \left(2 k_F - q\right)   < \w < v_F \left(2 k_F + q\right).
\end{equation}
 electron-hole pairs excitations \cite{graphenePlasmonicsBook} demonstrated in Fig. \ref{fig:dispersionGr} by shaded areas.

Substituting the long wavelength limit expression (\ref{eq:k+q}) in the overlap function (\ref{eq:overlapF}), the latter  reads
\begin{equation}
\label{eq:overlapF2}
F_{ss'}(\kk,\kk+\qq)=\begin{cases} ~1-\frac{q^2 }{4k^2}\sin^2\theta &\simeq 1  ~ ~\quad s=s' \text{ (intraband)}\\	
~ \frac{q^2}{4 k^2}\sin^2\theta &\simeq 0  ~\quad~ s \neq s' \text{ (interband)}
\end{cases}
\end{equation}
Equation (\ref{eq:overlapF2}) states that in  long wavelength limit the interband contribution can be neglected \cite{prb34_1986}, hence,  the Lindhard formula (\ref{eq:LindhardGr}) is simplified to
\begin{equation}
\label{eq:LindhardGr2}
\chi_0(\qq\rightarrow 0,\w)=-\frac{4}{V} \sum_\kk \left\lbrace \frac{f(\enrG{\kk+\qq}{+})-f(\enrG{\kk}{+}) }{\hbar z-\left(\enrG{\kk+\qq}{+} -\enrG{\kk}{+}\right)}+\frac{f(\enrG{\kk+\qq}{-})-f(\enrG{\kk}{-}) }{\hbar z-\left(\enrG{\kk+\qq}{-} -\enrG{\kk}{-}\right)}   \right\rbrace.
\end{equation}
 As it has been already mentioned, in  zero temperature limit, the Fermi-Dirac distribution $f(\enrG{\kk}{\pm})$ is simplified to Heaviside step function $\Theta(k_F\mp|\kk|)$. In this limit, the second term in the right hand of equation (\ref{eq:LindhardGr2}) is always zero, since $\Theta(k_F+|\kk|)=\Theta(k_F + |\kk+\qq|)=1$, which reflects that all states in the valence band are occupied. 
Making again the elementary transformation $\kk+\qq\rightarrow-\kk$ in the term of (\ref{eq:LindhardGr2}) that includes  $ f(\enrG{\kk+\qq}{+})$, we obtain
\begin{equation}
\label{eq:LindhardGr4} 
\chi_0(\qq\rightarrow 0,\w)= \frac{ 8 }{V} \sum_{ |\kk|<k_F  }  \frac{ \enrG{\kk+\qq}{+} -\enrG{\kk}{+} }{\left(\hbar z\right)^2-\left(\enrG{\kk+\qq}{+} -\enrG{\kk}{+}\right)^2}.
\end{equation}
Turning the summation (\ref{eq:LindhardGr4}) into integral, we read
\begin{equation}
\label{eq:IntegraGr} 
\chi_0(\qq\rightarrow 0,\w)= \frac{ 8}{(2\pi)^2} \int d^2|\kk|  \frac{ \enrG{\kk+\qq}{+} -\enrG{\kk}{+} }{\left(\hbar z\right)^2-\left(\enrG{\kk+\qq}{+} -\enrG{\kk}{+}\right)^2}.
\end{equation}
 Transforming to polar coordinates for $r=|\kk|$ and using the relation (\ref{eq:Grde}), we obtain the integral
\begin{equation}
\label{eq:IntegraGr2} 
\chi_0(\qq,\w)= \frac{ 2  E_F k_F q}{\pi^2\hbar^2\w^2 } \int_0^{1} dx \int_{0}^{2\pi} \frac{x \cos\theta+\frac{q}{2 k_F }\sin^2\theta }{1-\left(\frac{v_F q}{\w} \right)^2  \left( \cos\theta+\frac{q}{2 k_F x}\sin^2\theta \right) ^2}~ d\theta,
\end{equation}
where  $x=r/k_F$, $q=|\qq|$ and $\eta=0\Rightarrow z=\w$. In  non-static $(\w\gg v_F q)$ and  long wavelength ($q\ll k_F$) limits, we  expand the integrator of (\ref{eq:IntegraGr}) in series of $q$. Keeping up to first power of $q/k_F$, we obtain
\begin{equation}
\label{eq:IntegraGr3} 
\chi_0(\qq\rightarrow 0,\w)= \frac{ 2  E_F k_F q}{\pi^2\hbar^2\w^2} \int_0^{1} dx \int_{0}^{2\pi} \left( x \cos\theta +\frac{q}{2 k_F}\sin^2\theta \right)  d\theta.
\end{equation}
The evaluation of the integral (\ref{eq:IntegraGr3}) is trivial and leads to the polarizability function of graphene 
\begin{equation}
\label{eq:chiGr} 
\chi_0(\qq\rightarrow 0,\w)= \frac{   E_F }{\pi\hbar^2}\frac{q^2}{\w^2}.
\end{equation}
Using the RPA formula (\ref{eq:RPAeps}), we obtain the long wavelength dielectric function of graphene 
\begin{equation}
\label{eq:epsGr}
\varepsilon(q,\w)= 1 - \frac{2 e^2 E_F}{\hbar^2 \w^2}q
\end{equation}
indicating that at low energies doped graphene is described by a Drude type dielectric function with plasma frequency depends straightforward on doping amount, namely  the Fermi energy level $E_F$.  The plasma frequency of graphene monolayer is determined by the condition (\ref{eq:plasmonCondition}) and reads
\begin{equation}
\label{eq:plasmonGr}
\wplGr(q)= \sqrt{\frac{2 e^2 E_F }{\hbar^2}q}
\end{equation}
indicating the   $q^{1/2}$ dependence likewise  plasmons at a regular 2DEG. The most important  result  is  the presence of $\hbar$ in the denominator of equation (\ref{eq:plasmonGr}), which reveals that  plasmon in graphene are purely quantum modes, i.e. there is no classical plasmons in doped graphene. In addition, graphene plasmon frequency is proportional to $n^{1/4}$, which is different from classical 2D plasmon behavior where $\wplTwo \sim n^{1/2}$ \cite{prb75_2007,grapheneReview}.   This is a direct consequence of the quantum relativistic nature of graphene, since  Fermi energy  is defined differently in any case, namely,  $E_F\sim k_F \sim n^{1/2}$ in graphene, whereas, $E_F\sim k_F^2\sim n$ in 2DEG case.  
In Fig. \ref{fig:dispersionGr} we represent the plasmon dispersion relation in doped graphene.

\begin{SCfigure}
  \centering
\includegraphics[scale=.25]{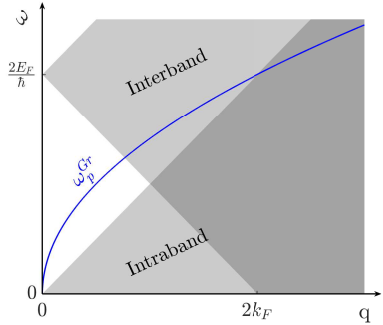}
  \caption{Blue solid line indicates the dispersion relation of graphene plasmons $(\wplGr)$. The shaded regimes represents the intra- and interband Landau damping where  plasmon decay to electron-hole pairs excitation.} \label{fig:dispersionGr}
\end{SCfigure}

\subsection{Graphene Plasmonic Metamaterial}

Multilayers of plasmonic materials  have been used for designing metamaterials provide electromagnetic propagation behavior not found under normal circumstances like negative refraction and epsilon-near-zero (ENZ) \cite{mariosENZ, prl109_2012, wang2}. {The bottleneck in creating plasmonic devices with any desirable characteristic has been the limitations of typical 3D  solids in producing perfect interfaces for the confinement of electrons and the features of dielectric host. This may no longer be a critical issue. The advent of truly two-dimensional  materials like graphene (a metal), transition-metal dichalcogenides (TMDC’s, semiconductors), and hexagonal boron nitride (hBN, an insulator) make it possible to produce structures with atomic-level control of features in the direction perpendicular to the stacked layers \cite{mariosENZ,jop15_2013}. This is ushering a new era in manipulating the properties of plasmons and designing devices with extraordinary behavior. } 

 Here, we propose a systematic method for constructing epsilon-near-zero (ENZ) metamaterials by appropriate combination on 2D materials. The aforementioned metamaterials exhibit interesting properties like diffractionless EM wave propagation with no phase delay \cite{mariosENZ}. We show analytically that EM wave propagation through layered heterostructures can be tuned dynamically by controlling the operating frequency and the doping level of the 2D metallic layers. Specifically, we  find that  multilayers of a plasmonic 2D material embedded in a dielectric host exhibit a plasmonic Dirac point (PDP), namely a point in wave-number space where two linear coexisting dispersion curves cross each other, which, in turn, leads to an effective ENZ behavior \cite{mariosENZ}.
To prove the feasibility of this design, we investigate numerically EM wave propagation in periodic plasmonic structures consisting of 2D metallic layers lying on $yz$ plane in the form of graphene, arranged periodically along the $x$ axis and possessing surface conductivity $\sigma_s$. The layers are embedded in a uniaxial dielectric host in the form of TMDC or hBN multilayers of thickness $d$ and with uniaxial relative permittivity tensor ${\Bar{\bar{\e}}}_d$  with diagonal components $\e_x \neq \e_y=\e_z$. We explore the resulting linear, elliptical, and hyperbolic EM dispersion relations which produce  ENZ effect, ordinary  and negative diffraction, respectively.

We solve the analytical problem under  TM polarization,  with the magnetic field parallel to the $y$ direction which implies that there is no interaction of the electric field with $\varepsilon_y$.  We consider a magnetically inert (relative permeability $\mu=1$) lossless host ($\e_x,\e_z\in\mathbb{R}$). For  monochromatic harmonic  waves in time the  Maxwell equations lead to  three equations  connecting the components of the $\EE$ and $\HH$ fields. For the longitudinal component \cite{mariosENZ, prl109_2012},  $E_z=(\im\eta_0/k_0\e_z)(\partial H_y/\partial x)$ where  $\eta_0=\sqrt{\mu_0/\varepsilon_0}$ is the free space impedance. Defining the vector of the transversal field components as  $\vPsi=\left(E_x ,\ H_y\right)^T$, gives \cite{mariosENZ} 
\begin{equation}
\label{eq:FullMatrix}
\im\frac{\partial}{\partial z}\vPsi= k_0\eta_0\left({\begin{array}{cc} 0    & 1+\frac{1}{k_0^2}\frac{\partial}{\partial_x}\frac{1}{\e_z}\frac{\partial}{\partial_x}\\ 
               \frac{\e_x}{\eta_0^2} & 0\      \end{array}} \right) \vPsi
\end{equation}
Assuming EM waves propagating along the $z$ axis, viz. $\vPsi(x,z)= \vPsi(x) e^{\im k_z z}$, equation (\ref{eq:FullMatrix}) leads to an eigenvalue problem for the  wavenumber $k_z$ of the plasmons along $z$ \cite{mariosENZ, prl109_2012}. The metallic 2D planes are assumed to carry a surface current $J_s=\sigma_s E_z$, which acts as a boundary condition in the eigenvalue problem. Furthermore, infinite number of 2D metals are considered to be arranged periodically, along $x$ axis, with structural period $d$.
The magnetic field reads $H_y^-(x) e^{\im k_z z}$ for $-d<x<0$  and $H_y^+(x) e^{\im k_z z}$ for $0<x<d$ on either side of the metallic plane at $x=0$,  with boundary conditions  $H_y^+(0)-H_y^-(0)=\sigma_s E_z(0)$ and $\partial_xH_y^+(0)=\partial_xH_y^-(0)$.
Due to the periodicity we use Bloch theorem along $x$ as $H_y^+(x)=H_y^-(x-d)e^{\im k_x d}$,  with Bloch wavenumber $k_x$. As a result,  we arrive at the dispersion 
relation \cite{mariosENZ, prl109_2012, wang2}:
\begin{equation}
\label{eq:dispersion}
F(k_x,k_z) =  \cos(k_x d) - \cosh(\kappa d) + \frac{\xi\kappa}{2}\sinh(\kappa d) = 0
\end{equation}
where $\kappa^2=(\e_z/\e_x)(k_z^2-k_0^2\e_x)$  expresses the anisotropy of the  host medium and  $\xi=-(\im\sigma_s\eta_0/k_0\e_z)$  coincides with the so-called ``plasmonic thickness''  which determines the SPP decay length \cite{mariosENZ, prl109_2012, wang2} {In particular, $\xi$ is twice the SPP penetration length and defines the maximum distance between two metallic layers where the plasmons are strongly interacting \cite{mariosENZ, prl109_2012, wang2}.}  We point out that for lossless 2D metallic planes $\sigma_s$ is purely imaginary and $\xi$ is purely real (for $\e_z\in\mathbb{R}$).  At the center of the first Brillouin zone $k_x=0$, the equation has a trivial solution \cite{prl109_2012} for $\kappa=0 \Rightarrow k_z=\DP$ which corresponds to propagation of $x$-polarized fields  travelling into the host medium with refractive index $\sqrt{\e_x}$  without interacting with the 2D planes which are positioned along $z$ axis \cite{MyNJP16_2014}.  Near the Brillouin zone center $(k_x/k_0\ll 1$ and $\kappa \simeq 0)$   and under the assumption of a very dense grid $(d\rightarrow 0)$,  we obtain $k_x d\ll 1$  and  $\kappa d\ll 1$,  we Taylor expand the dispersion equation (\ref{eq:dispersion}) to second order in $d$, hence
\begin{equation}
\label{eq:taylor1}
\frac{k_z^2}{\e_x}+\frac{d}{(d-\xi)\e_z}k_x^2=k_0^2.
\end{equation}
The approximate relation (\ref{eq:taylor1})   is identical to that of an equivalent homogenized medium described by dispersion: $k_z^2/\e^{\rm eff}_{x}+k_x^2/\e^{\rm eff}_{z}=k_0^2$ \cite{mariosENZ,jop15_2013}. Subsequently, from a metamaterial point of view, the entire system is treated as a homogeneous anisotropic medium with effective relative permittivities  given by
\begin{equation}
\label{eq:effperms}
\e^{\rm eff}_{x}=\e_x\ , \; \;  
\e^{\rm eff}_{z}=\e_z+\im \frac{\eta_0 \sigma_s}{k_0d}=\e_z\frac{d-\xi}{d}.
\end{equation}
We read from equation (\ref{eq:effperms})  the capability to control the behavior of the overall structure along the $z$ direction. For instance the choice $d=\varepsilon_z/(\varepsilon_z-\varepsilon_x)\xi$ leads to an isotropic effective medium with $\e^{\rm eff}_{z}= \e^{\rm eff}_{x}$ \cite{mariosENZ}.

For the lossless case ($\IM{\xi}=0$), we identify two  interesting regimes, viz. the strong plasmon coupling  for $d<\xi$ and the weak plasmon coupling  for $d>\xi$. In both cases plasmons develop along $z$ direction at the interfaces between the conducting planes and the dielectric host. In the strong coupling case ($d<\xi)$, plasmons of adjacent interfaces interact strongly each other. As a consequence,  the shape of the supported band  of equation (\ref{eq:taylor1}), in the $(k_x,k_z)$ plane, is hyperbolic  (dashed red line in Fig. \ref{fig:maps}(a))and the system behaves as a hyperbolic metamaterial \cite{mariosENZ, prl109_2012, MyNJP16_2014} with  $\e^{\rm eff}_{x}>0$, $\e^{\rm eff}_{z}<0$. On the other hand, in the weak plasmon coupling  $(d>\xi)$, the interaction between plasmons of adjacent planes is very weak. In this case the shape of the dispersion relation  (\ref{eq:taylor1}) on the $(k_x,k_z)$ plane is an ellipse (dotted black line in Fig. \ref{fig:maps}(a)) and the systems acts as an ordinary anisotropic media  with $\e^{\rm eff}_{z}, \e^{\rm eff}_{x}>0$ \cite{mariosENZ}. We note that in the case $\xi<0$ the system does not support plasmons and the supported bands are always ellipses \cite{mariosENZ}.   When either the 2D medium (${\rm Re}[\sigma_s]\ne 0$)  or the host material are lossy,  a similar separation holds by replacing $\xi$ by ${\rm Re}[\xi]$.

The most interesting case is the linear dispersion,  where $k_z$ is linearly dependent on $k_x$ and $dk_x/dk_z$ is constant for a wide range of  $k_z$ \cite{mariosENZ, prl109_2012}.  When this condition holds,  the spatial harmonics travel with the same group velocity into the effective medium   \cite{mariosENZ, prl109_2012}.  To engineer our structure to exhibit a close-to-linear dispersion relation,  we inspect the approximate version of equation (\ref{eq:taylor1}): a huge coefficient for $k_x$ will make $k_0^2$ on the right hand side  insignificant; if $\xi=d$, the term proportional to $k_x^2$  increases without bound yielding a linear relation between $k_z$ and $k_x$. With this choice, $\sigma_s=-\im(k_0d\e_z/\eta_0)$,  and substituting in the exact dispersion relation Eq. (\ref{eq:dispersion}),  we find that $(k_x,k_z)=(0,k_0\sqrt{\e_x})$ becomes a saddle point for the transcendental function $F(k_x,k_z)$ giving rise to the conditions for the appearance of two permitted bands, namely two lines on the $(k_x,k_z)$ plane across which $F(k_x,k_z)=0$.  This argument connects a mathematical feature,  the saddle point of the dispersion relation, with a physical feature,  the crossing point of the two coexisting linear  dispersion curves,  the Plasmonic Dirac point \cite{mariosENZ}  (solid blue line in Fig. \ref{fig:maps}(a)).  From a macroscopic point of view, the choice $\xi=d$ makes  the effective permittivity along the $z$ direction vanish, as is evident from equation (\ref{eq:effperms}). As a result, the existence of a  PDP makes  the effective medium behave like an ENZ material in one direction ($\e^{\rm eff}_{z}=0$). 

 The plasmonic length $\xi$ is, typically, restricted in few nanometers ($\xi <100$nm).    Regular dielectrics always present imperfections in nanoscales, hence, the use of regular materials as dielectric hosts are impractical. Furthermore, graphene usually exfoliates or grows up on other 2D materials. Because of the aforementioned reasons, it is strongly recommended the dielectric host to be   also a 2D material with atomic scale control of the thickness $d$ and no roughness.
For instance, one could build  a dielectric host by stacking 2D layers of materials   molybdenum disulphide (MoS$_2$)  \cite{rodrickPRB} with essentially perfect planarity,  complementing the planarity of graphene.

Substituting the graphene dielectric function (\ref{eq:epsGr}) into the formula (\ref{eq:sigmaEps}), we calculate the two-dimensional Drude type conductivity of graphene  \cite{jablanPRB,prl109_2012,jop15_2013}
\begin{equation}
\label{eq:drudeGr}
\sigma_s(\omega)=\frac{\im  e^2 \mu_c}{\pi\hbar^2(\omega+\im/\tau)},
\end{equation} 
where $\mu_c$ is the tunable chemical potential equal to Fermi energy $E_F$ and {$\tau$ is the transport  scattering time of the electrons \cite{prl109_2012,  jablanPRB} introduced in the same manner as in equation (\ref{eq:RPAresult3Db}).} In what follows, we use bulk MoS$_2$, which at THz frequencies  is assumed lossless with a diagonal permittivity tensor of elements, $\e_x\cong 3.5$ (out of plane)  and $\e_y=\e_z\cong13$ (in plane) \cite{rodrickPRB}.  

The optical losses of graphene are taken into account using $\tau=0.5$ ps \cite{prl109_2012}. Since the optical properties of the under investigated system can be controlled  by tuning the doping amount, the operating frequency or the structural period, in  {Fig. \ref{fig:maps}(b) we show proper combinations of  $\mu_c$   and  operational wavelength in free space  $\lambda$ which lead to a PDP  for several values of lattice density distances ($d={\rm Re}[\xi]$ in $\rm nm$) \cite{mariosENZ}.}
To illustrate, for a reasonable  distance between successive graphene planes of $d=20~\rm nm$,  the real (Fig. \ref{fig:maps}(c)) and imaginary (Fig. \ref{fig:maps}(d)) effective permittivity values that can be emulated by this specific {graphene-MoS$_2$}  architecture determine the device characteristics at different frequencies  and graphene doping levels. Positive values of ${\rm Re}[\e^{\rm eff}_{z}]$  are relatively moderate and occur  for larger frequencies and lower doping levels of graphene; on the other hand, ${\rm Im}[\e^{\rm eff}_{z}]$ is relatively small in the ENZ region  as indicated by a dashed line in both graphs \cite{mariosENZ}.  On the other hand, losses become larger as ${\rm Re}[\e^{\rm eff}_{z}]$  gets more negative.  

\begin{figure}[ht!]
\centering 
\includegraphics[scale=0.25]{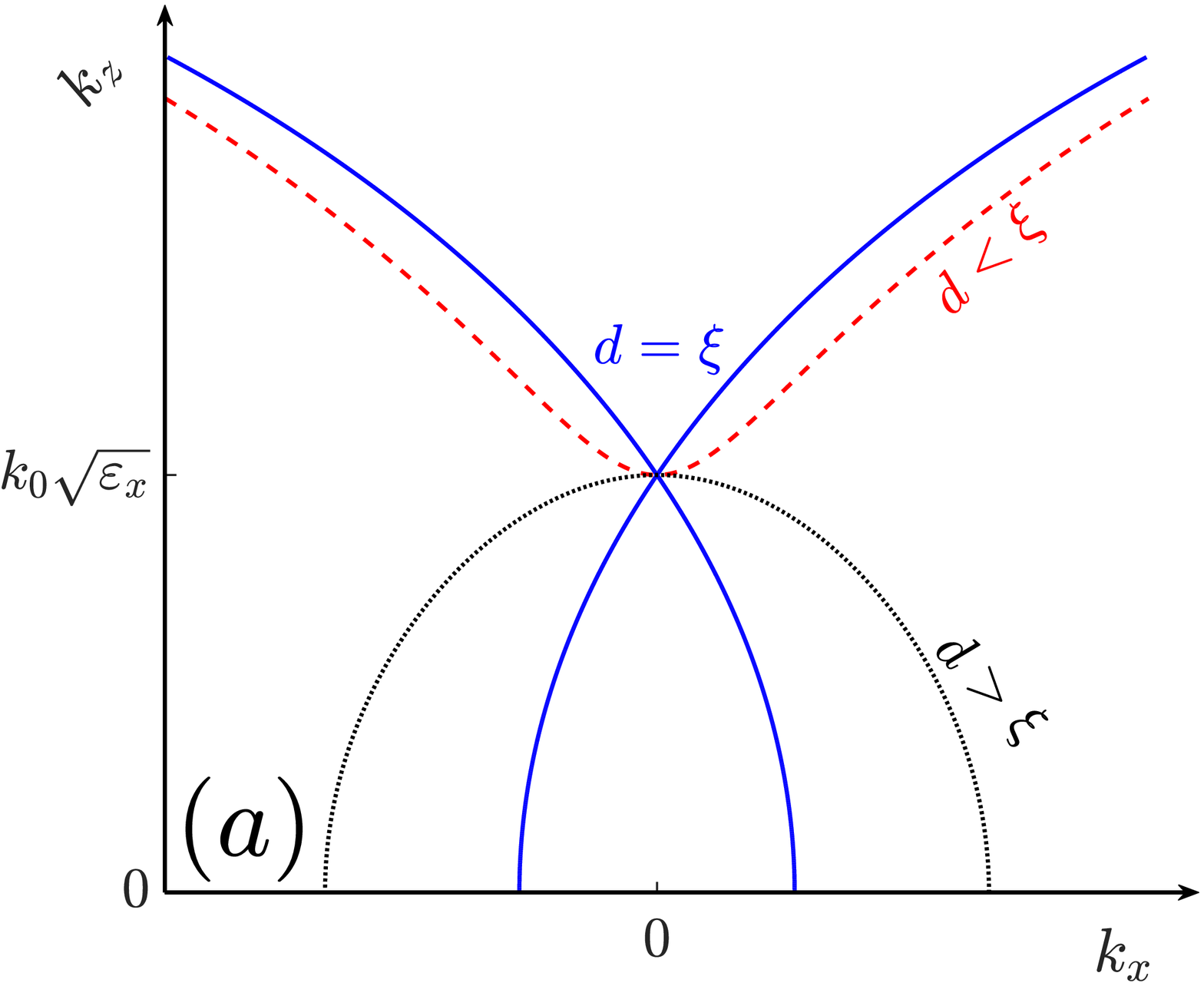} 
\includegraphics[scale=0.25]{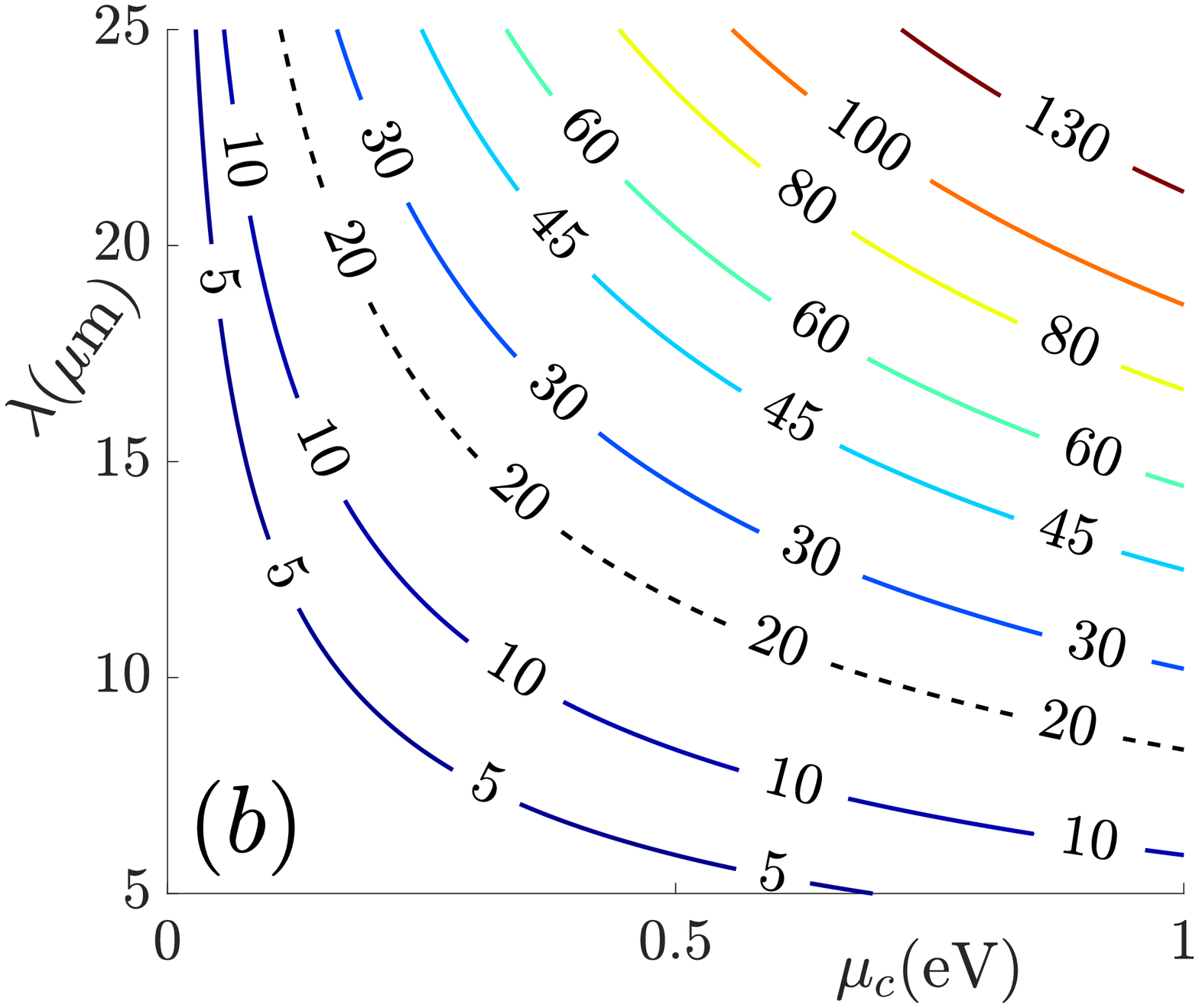}\\ 
\includegraphics[scale=0.25]{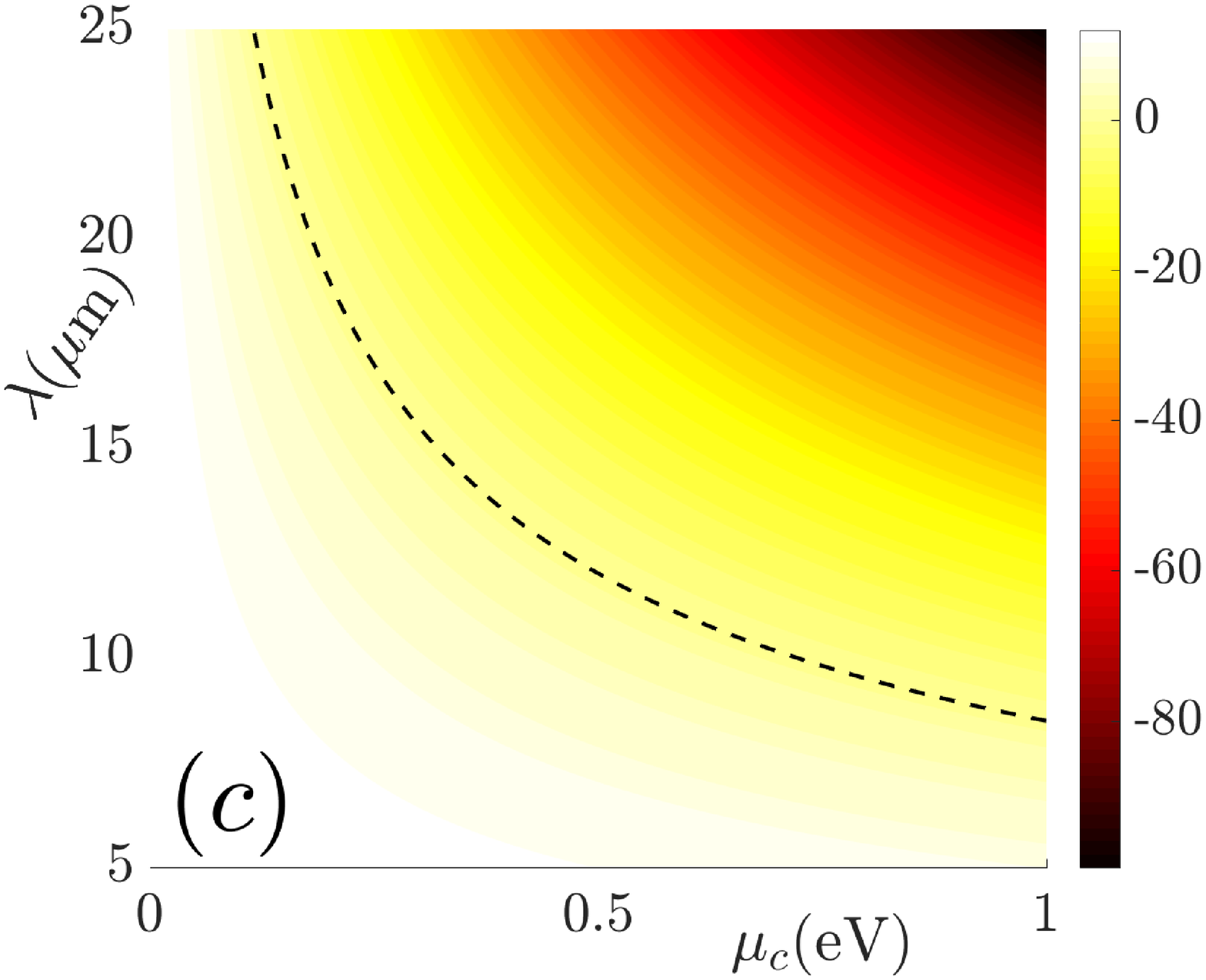}  
\includegraphics[scale=0.25]{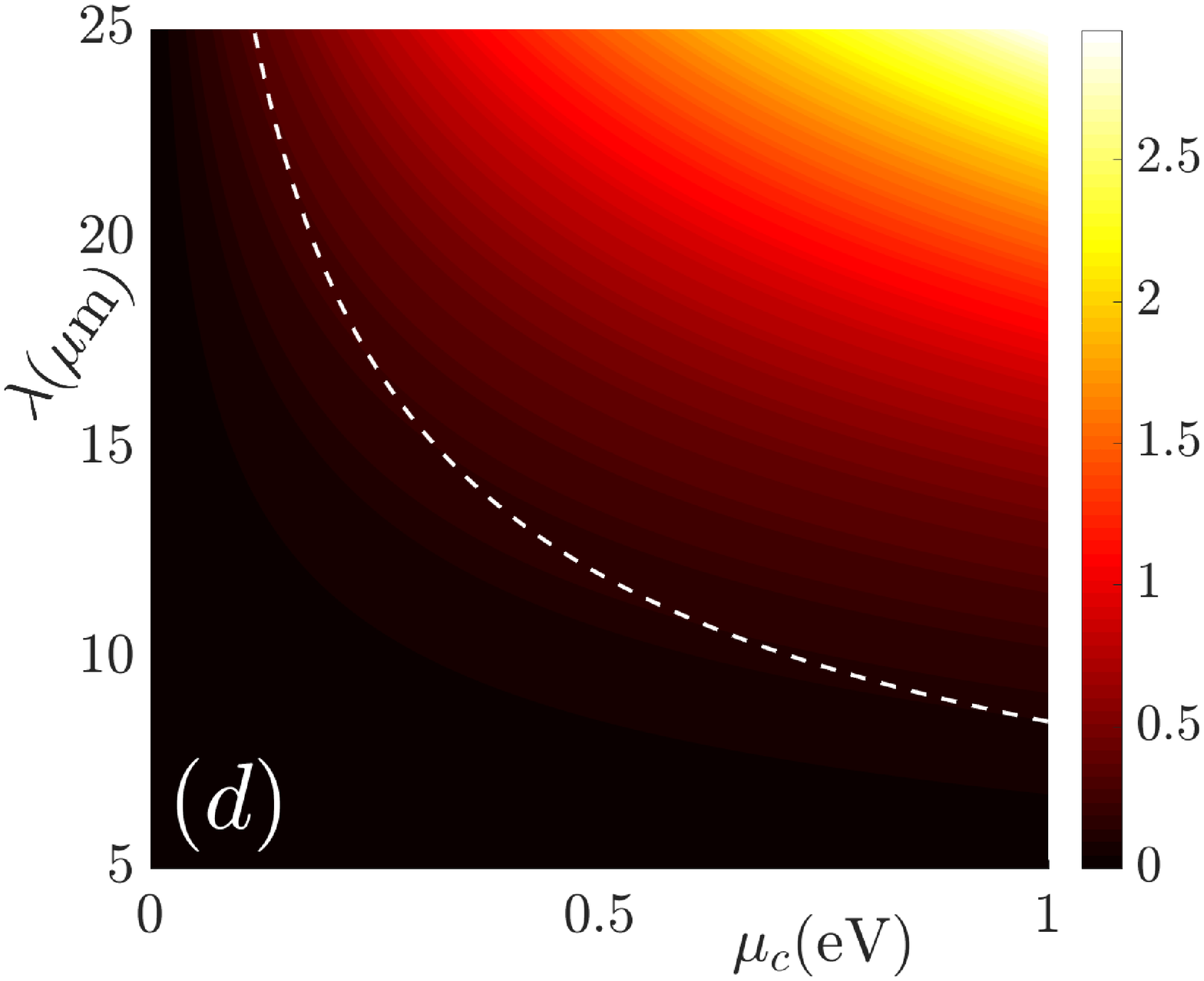}  
\caption{(a) The three supported dispersion plasmonic bands in $(k_x, k_z)$ plane: hyberbolic (dashed red), elliptical (dotted black) and linear (solid blue) where plasmonic Dirac point (PDP) appears. (b) Combinations of graphene doping $\mu_c$  and free space operational wavelengths $\lambda$  leading to epsilon-near-zero (ENZ) behavior (PDP in dispersion relation)   for several lattice periods $d$ (in $\rm{nm}$).    (c) Real (d) imaginary parts of the effective permittivity $\e_z^{\rm eff}$    for the choice $d=20~\rm nm$ (dashed line in (b)); dashed curves indicate the ENZ regime.}
\label{fig:maps}
\end{figure}

To examine the actual electromagnetic field distribution in our graphene-MoS$_2$  configuration we simulate the EM wave propagation through  two  finite structures  consisting of 40 and 100 graphene planes with ${\rm Re}[\xi]=20.8~\rm nm$ and for operational wavelength  in vacuum $\lambda=12~\rm \mu m$ ($f=25$ THz$=0.1$ eV). In order to have a complete picture of the propagation properties, we excite the under investigating structures with a 2D dipole magnetic source as well as with a TM plane wave source. In particular, the 40-layered structure is excited by a 2D magnetic dipole source, which is positioned close to one of its two interfaces  and oriented parallel to them, denoted by a white dot in Figs. \ref{fig:dispersionNUM} (a)-(c). On the other hand, the 100-layered configuration is excited by a plane source which is located below the multilayer and is rotated by $20^o$ with respect to the interface; the blue arrow in Fig. \ref{fig:dispersionNUM}(d) indicates the direction of the incident wave.  The normalized to one spatial distribution of the magnetic field value is shown in Fig. \ref{fig:dispersionNUM} in color representation, where the volume containing the graphene multilayers is between the dashed blue lines.  To minimize the reflections the background region is filled  with a medium of the same dielectric properties as {MoS$_2$}.  In Figs. \ref{fig:dispersionNUM}(a)(d), the system is in the critical case ($d={\rm Re}[\xi]$), where the waves propagate through the graphene sheets without dispersion as in an ENZ medium.  In Figs. \ref{fig:dispersionNUM}(b)(e) the interlayer distance is $d=0.7{\rm Re}[\xi]$ (strong plasmon coupling regime) and the system shows negative (anomalous) diffraction.  In Fig. \ref{fig:dispersionNUM}(c)(f) $d=1.5{\rm Re}[\xi]$ (weak plasmon coupling regime) and the EM wave shows ordinary diffraction through the graphene planes \cite{mariosENZ}.

\begin{figure}[ht!]
\centering
\includegraphics[scale =0.17]{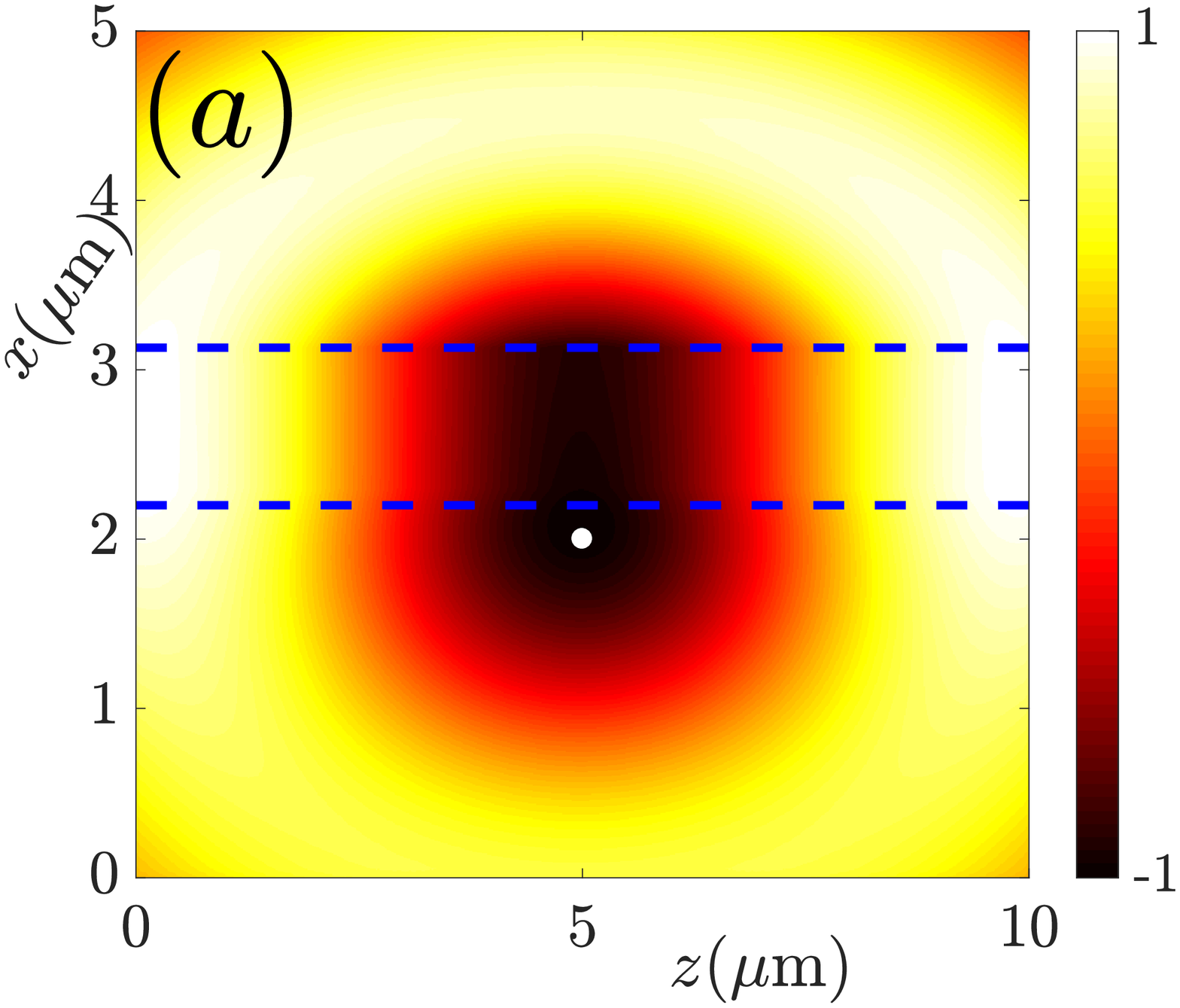} 
\includegraphics[scale =0.17]{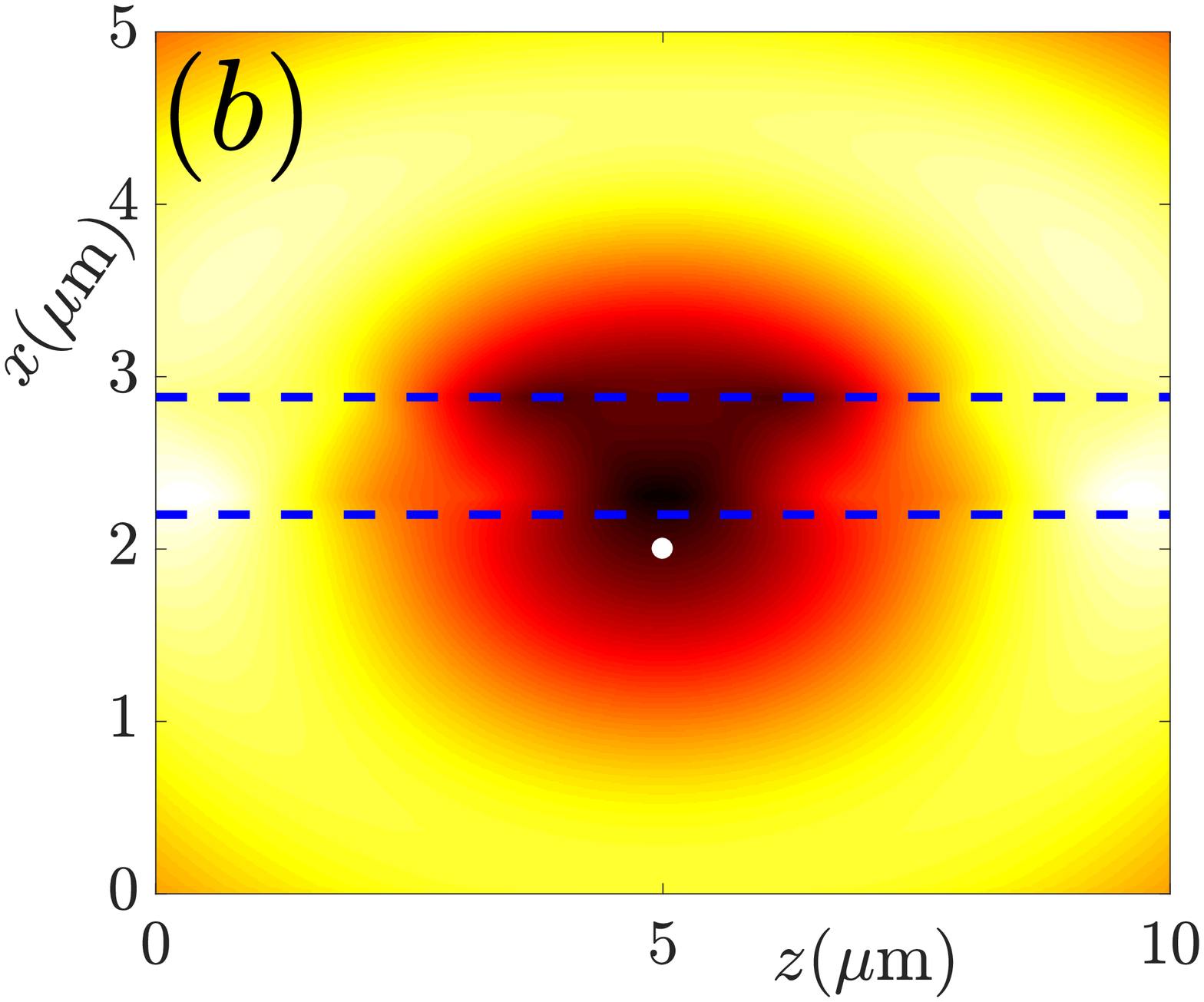} 
\includegraphics[scale =0.17]{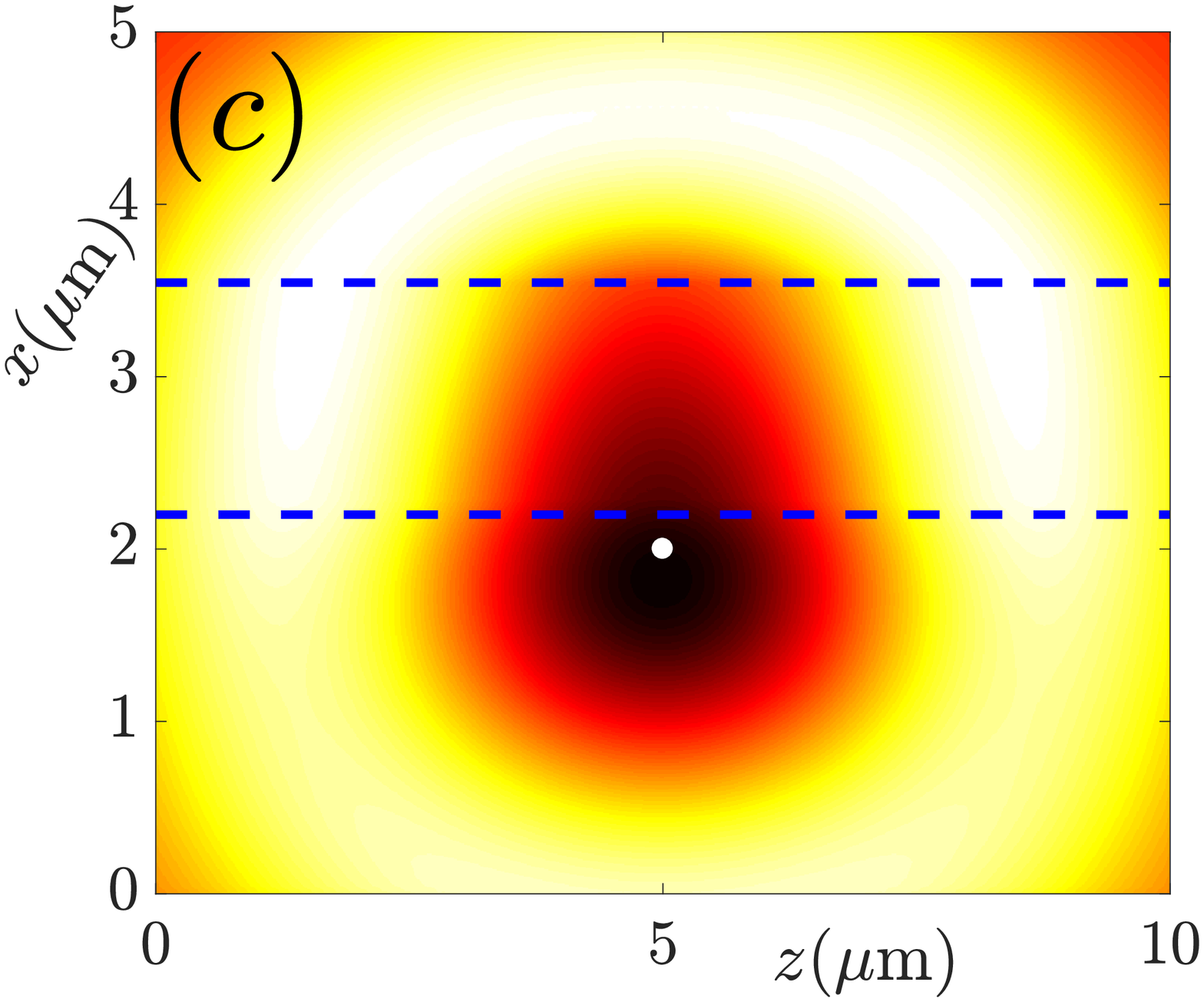}\\
\includegraphics[scale =0.17]{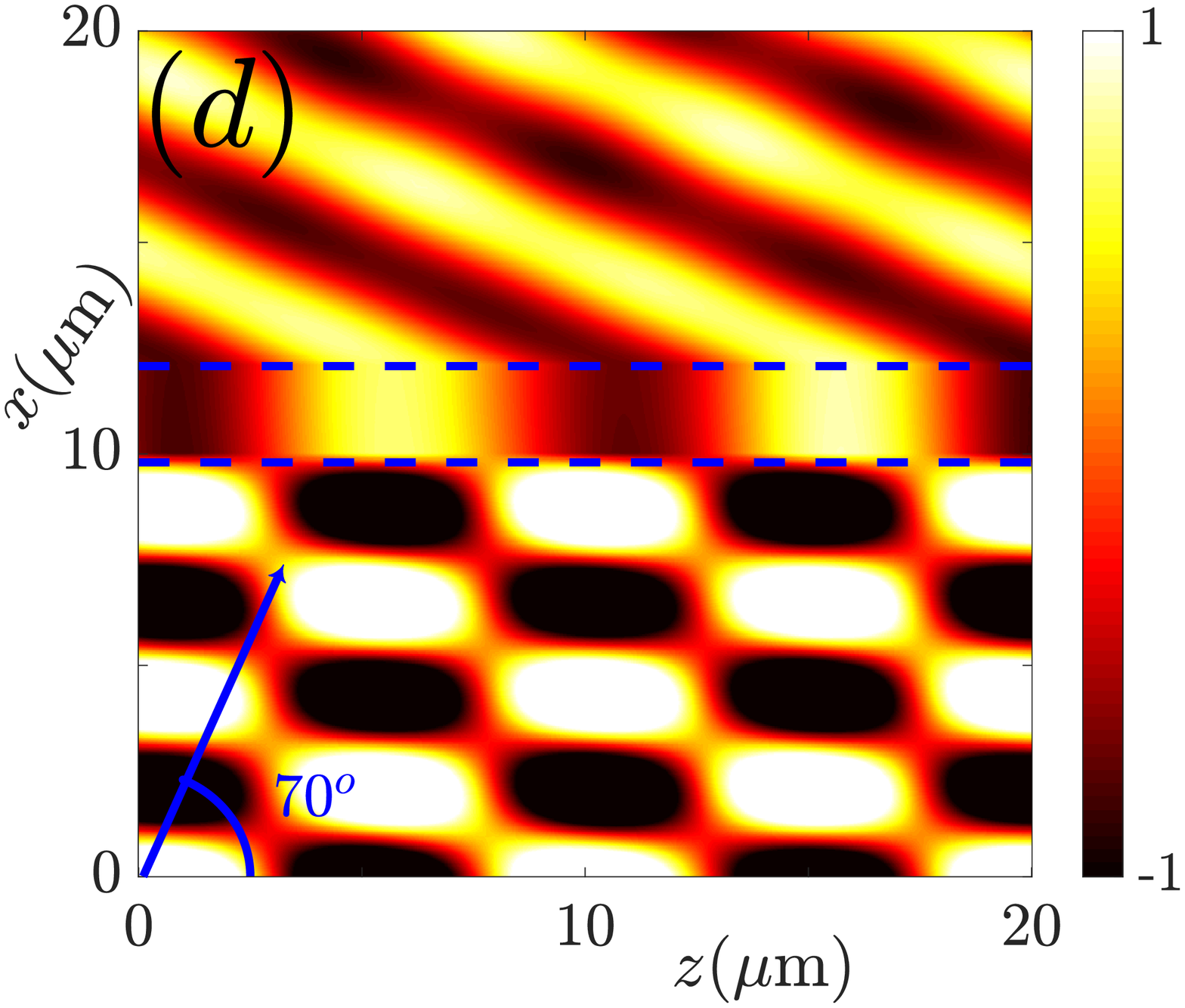} 
\includegraphics[scale =0.17]{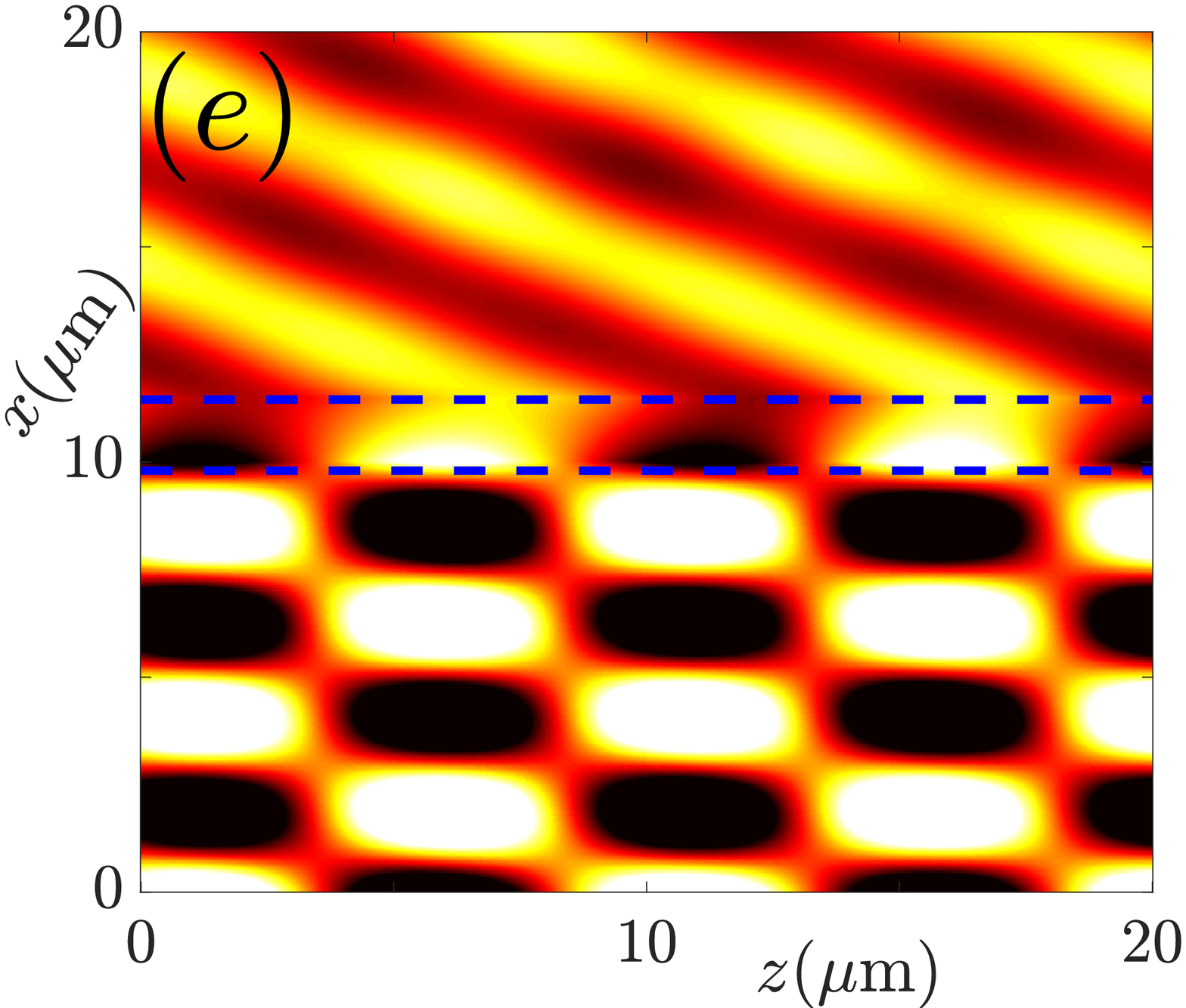} 
\includegraphics[scale =0.17]{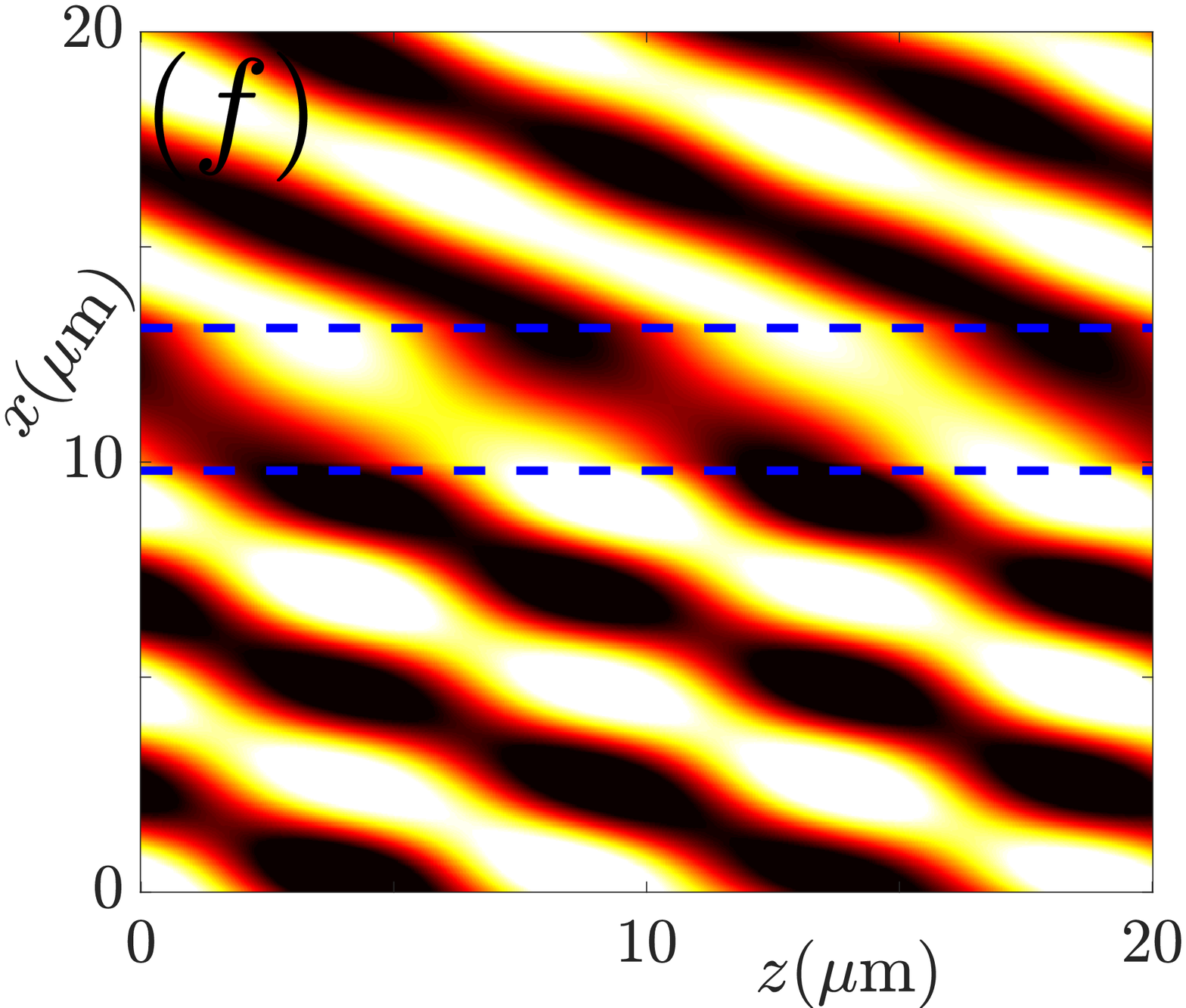}

\caption{Spatial distribution of the magnetic field (colormap) of  graphene-MoS$_2$ multilayer structure located between the blue dashed lines and embedded in MoS$_2$ background. In (a)-(c) the metamaterial consists of 40 graphene sheets and excited by  a magnetic dipole (white dot). In (d)-(f) the structure is comprised by 100 graphene layers and excited by a TM plane wave source located at $y=0$ and rotated $20^o$ with respect to the interface. (a)(d) $d={\rm Re}[\xi]$ (ENZ behavior).  (b)(e) $d=0.7{\rm Re}[\xi]$ hyperbolic metamaterial. (c)(f) $d=1.5{\rm Re}[\xi]$ elliptical medium, where ${\rm Re}[\xi]=20.8~\rm nm$. Due to  high reflections in (d)(e), we observe pattern formation of stationary waves below the metamaterial.}
\label{fig:dispersionNUM}
\end{figure}

\section{Conclusion}
In summary, we have studied volume and surface plasmons beyond the classical plasma model. In particular, we have described electronic excitations in solids, such as plasmons and their damping mechanism, viz. electron-hole pairs excitation, in the context of the quantum approach random phase approximation (RPA); a powerful self consistent theory for determining the dielectric function of solids including screening non-local effect. The dielectric function and, in turn, the plasmon dispersion relation have been calculated for a bulk metal, a two-dimensional electron gas (2DEG) and for graphene, the famous two-dimensional semi-metal. The completely different dispersion relation between plasmon in three- and two-dimensional metals has been pointed out. Furthermore, we have highlighted the fundamental difference between plasmons in a regular 2DEG and in doped graphene, indicating that plasmons in graphene are purely quantum modes, in contrast to plasmons in 2DEG which originate from classical laws. Moreover, the propagation properties of surface plasmon polariton (SPP), a guided collective oscillation mode, have been also investigated.  For the completeness of  our theoretical investigation, we have outlined two applications. Firstly, we have examined SPPs properties along an interface consists of a bulk metal and an active (gain) dielectric. We have found that there is a gain value for which the metallic losses have been completely eliminated resulting to lossless SPP propagation. Secondly, we have investigated a plasmonic metamaterial composed of doped graphene monolayers. We have shown that depending on operating frequency, doping amount and interlayer distance between adjacent graphene layers, the wave propagation properties presents epsilon-near-zero behavior, normal and negative refraction, providing a metamaterial with tunable optical properties.

\section*{Acknowledgements}
We acknowledge discussions with D. Massatt and E. Manousakis and partial support by the European Union under programs  H2020-MSCA-RISE-2015-691209-NHQWAVE
and  by the Seventh Framework Programme (FP7-REGPOT-
2012-2013-1) under grant agreement no 316165.  We also acknowledge support by EFRI 2-DARE NSF Grant No. 1542807 (M.M); ARO MURI Award No. W911NF14-0247 (E.K.). We used computational resources on the Odyssey cluster of the FAS Research Computing Group at Harvard University.

\end{document}